\documentclass[prb,twocolumn,showpacs,preprintnumbers,amsmath,amssymb]{revtex4}

\usepackage{graphicx}
\graphicspath{{fig/}}
\usepackage{dcolumn}
\usepackage{bm}
\usepackage{amsmath,amssymb}
\usepackage[T1]{fontenc}
\usepackage{textcomp}
\usepackage{color}

\begin{document}

\title{Magnon transport through microwave pumping
}

\author{Kouki Nakata,$^{1}$ Pascal Simon,$^2$ and Daniel Loss$^{1}$}

\affiliation{$^1$Department of Physics, University of Basel, Klingelbergstrasse 82, CH-4056 Basel, Switzerland   \\
$^2$Laboratoire de Physique des Solides, CNRS UMR-8502, Universit$\acute{e} $ Paris Sud, 91405 Orsay Cedex, France}

\date{\today}

\begin{abstract}
We present a microscopic theory of magnon transport in ferromagnetic insulators (FIs). 
Using  magnon injection through microwave pumping, we propose a way to generate  magnon dc currents and show how to enhance their amplitudes in hybrid ferromagnetic insulating junctions. To this end focusing on a single FI, we first  revisit microwave pumping at finite (room) temperature from the microscopic viewpoint of  magnon injection. Next, we apply it to two kinds of hybrid ferromagnetic insulating junctions. The first is the junction between a quasi-equilibrium magnon condensate and magnons being pumped by microwave, while the second is the junction between such pumped magnons and noncondensed magnons. We show that quasi-equilibrium  magnon condensates generate  ac and dc magnon currents, while noncondensed magnons produce essentially a dc magnon current. The ferromagnetic resonance (FMR) drastically increases the density of the pumped magnons and enhances such magnon currents.
Lastly, using microwave pumping in a single FI, we discuss the possibility that a magnon current through an Aharonov-Casher phase flows persistently even at finite temperature. We show that such a magnon current arises even at finite temperature in the presence of magnon-magnon interactions. Due to FMR, its amplitude becomes much larger than the condensed magnon current. 
\end{abstract}


\maketitle

\section{Introduction}
\label{sec:intro}

Since the observation of spin-wave spin currents\cite{spinwave} in Y$_3$Fe$_5$O$_{12}$ (YIG), ferromagnetic insulators (FIs) have been playing a major  role in spintronics\cite{awschalom,mod}.
 All the fascinating features of FIs arise from magnons\cite{demokritov,ultrahot,MagnonSupercurrent,KKPD,magnon2,Kevin2,takei,bender,Kopietz,morimae,MagnonQubit,2dMagnonLifetime} (i.e., spin-waves), which are bosonic low-energy collective excitations (i.e., quasiparticles) in a magnetically ordered spins system. The use of the magnon degrees of freedom for transport perspectives~\cite{magnon2} has led to a new field called magnonics\cite{magnonics,spincalreview}.
 
 \begin{figure}[h]
\begin{center}
\includegraphics[width=6.5cm,clip]{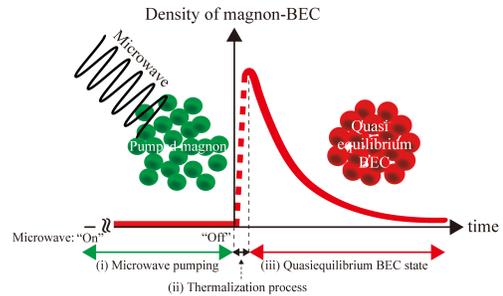}
\caption{(Color online)
A schema of the time-evolution of the quasi-equilibrium magnon-BEC density due to microwave pumping where
(i) microwave pumping, (ii) thermalization (or relaxation) process, and (iii) quasi-equilibrium magnon-BEC state. 
This schematic picture is based on the experiment by Serga {\textit{et al.}} of Ref. [\onlinecite{ultrahot}].
The applied microwave drives the system into a nonequilibrium steady state and continues to excite the zero-mode of magnons (i.e., pumped magnons).
After microwave is switched off, the system experiences a thermalization process for a short time, and immediately reach a quasi-equilibrium state where the pumped magnons can form a quasi-equilibrium magnon-BEC.
The thermalization period is\cite{ultrahot} about $50$ns, while the decay time of quasi-equilibrium magnon-BECs amounts\cite{ultrahot} to $400$ns and they can survive about for $1 \mu $s. 
\label{fig:BEC} }
\end{center}
\end{figure}
 
The main advantages of magnons in FIs can be summarized in the following  points:
First, transport of magnons in insulating magnets has a lower power absorption compared  to electronic conductors\cite{Trauzettel}.
Next, it has been shown experimentally by Kajiwara {\textit{et al.}}\cite{spinwave} that    magnon currents can carry spin-information over distances of several millimeters, much further than what is typically possible when using spin-polarized conduction electrons in magnetic metals. 
Third, magnons are quasi-particles (i.e., magnetic excitations) which are not conserved. Then, using microwave pumping\cite{demokritov,ultrahot,MagnonSupercurrent,ThermalizationHill,ISHE1,mod2}, they can be directly injected to FIs and their density controlled.  
A further advantage of magnons  arises from the feature that magnons are bosonic excitations and as such they can form a macroscopic coherent\cite{morimae} state, namely, a quasi-equilibrium Bose-Einstein condensate (BEC)\cite{He3Bunkov,bunkov,Yukalov,BatistaBEC}. Demokritov {\textit{et al.}}\cite{demokritov} have indeed experimentally shown that such a  magnon-BEC  can be generated even at room temperature\footnote{This proves that  low temperatures  are not required\cite{oshikawa}, which may be of crucial importance in view of applications.} in YIG by using microwave pumping, and Serga {\textit{et al.}}\cite{ultrahot} have recently addressed the relation between microwave pumping and the resulting {\it quasi-equilibrium}\footnote{Regarding the theoretical aspects of their quasi-equilibrium magnon-BECs, see Refs. [\onlinecite{bunkov,Yukalov,BatistaBEC,rezende,troncoso}] and also references therein.} magnon-BEC (Fig. \ref{fig:BEC}); the applied microwave drives the system into a {\it{nonequilibrium}} steady state\cite{bunkov,Yukalov} and continues to populate the zero-mode of magnons. After switching off the microwaves, the system undergoes a thermalization\cite{ThermalizationHill,demidov2,Vannucchi,Vannucchi2} (or relaxation) process and thereby reaches a quasi-equilibrium state where the pumped magnons form a quasi-equilibrium magnon-BEC\cite{MagnonSupercurrent}. This magnon-BEC is not the ground state but a metastable state\cite{BatistaBEC} that corresponds to a macroscopic coherent precession\cite{bunkov} in terms of spin variables which can last for about 1$\mu$s.

Motivated by these experimental progress,  we microscopically analyzed in Ref. [\onlinecite{KKPD}] the transport\cite{MagnonSupercurrent} properties in such\cite{demokritov,ultrahot} realized quasi-equilibrium magnon-BECs in FI junctions, and discussed the ac and dc Josephson effects through the Aharonov-Casher (A-C) phase\cite{casher}, being a special case of a Berry phase~\cite{Mignani,Hea} in this system. We  found that this A-C phase gives  a handle to electromagnetically control the Josephson effects and to realize a persistent\cite{LossPersistent,LossPersistent2,Kopietz} magnon-BEC current~\cite{bunkov,sonin,takei,sigrist} in a ring. An experimental proposal to directly measure this magnon-BEC current also has been theoretically proposed in Ref. [\onlinecite{KKPD}] through measurements of the induced electric field. 

Toward the direct measurement of such magnon currents, one concern\cite{KKPD} might be, 
however, that the amount of the magnon currents and the resulting electric fields (of the order of $\mu $V/m) is still\cite{magnon2,KKPD} very small. This might be an obstacle to the feasibility of such experiments at present.

\begin{figure}[h]
\begin{center}
\includegraphics[width=7cm,clip]{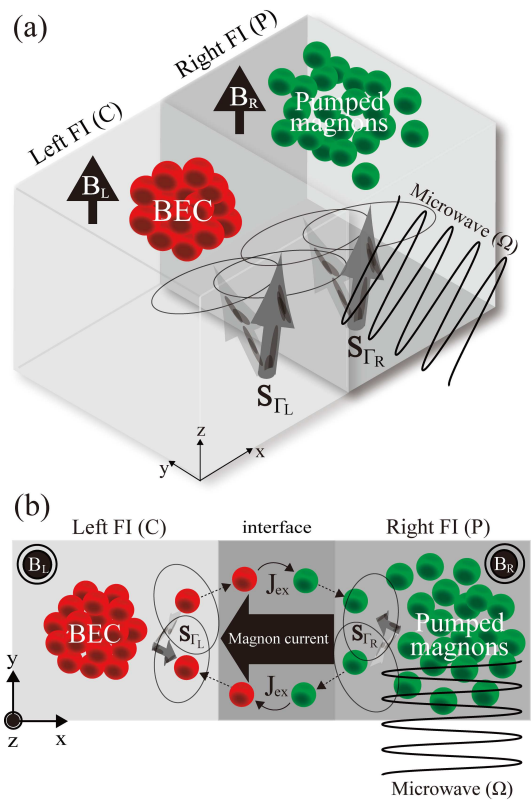}
\caption{(Color online)
Schematic representations of a C-P junction between two FIs, one prepared  as BEC (left FI)  and the other (right FI) with pumped magnons. 
(a) Both magnetic fields are applied along the $z$-axis, ${\bf B}_{\rm{L(R)}} = B_{\rm{L(R)}} {\bf e}_z$. The circles in the right FI are the magnons pumped by microwaves (frequency $\Omega \sim  100 $MHz)  and the cloud of circles in the BEC  represents the quasi-equilibrium magnon condensate. 
Both phases correspond to a macroscopic coherent precession in terms of spin variables ${\bf S}_{\Gamma_{\rm L}}$ and ${\bf S}_{\Gamma_{\rm R}}$. A distinct difference, however, is that the frequency of the magnon-BEC is controlled by the applied magnetic field $B_{\rm{L}} $ ($\sim $mT), while that of the pumped magnons by the frequency $\Omega $ of the applied microwave. The relation between $ g\mu_{\rm{B}}  {B}_{{\rm{L}}}/\hbar $ and $ \Omega $ plays the key role in the ac-dc conversion of magnon currents.
Tuning $\hbar \Omega = g\mu_{\rm{B}}  {B}_{{\rm{R}}} $, a FMR occurs in the right FI and consequently, the number of pumped magnons drastically increases.
(b) Close-up of the C-P junction. Only the boundary spins ${\bf S}_{\Gamma_{\rm L}}$ in the left FI and ${\bf S}_{\Gamma_{\rm{R}}}$ in the right FI are relevant to magnon transport. We assume a large spin $S \gg 1 $.
The two FIs are separated by an interface and thereby weakly exchange-coupled with strength $J_\textrm{ex}$. For typical values  $J_\textrm{ex}\sim  0.1 \mu $eV this gives rise to an interface of width $ \sim  10$\AA.
\label{fig:system} }
\end{center}
\end{figure}

To overcome this problem,  we propose to use instead  magnon injection through microwave pumping\cite{demokritov,ultrahot,ISHE1} which offers an alternative method to enhance magnon currents in FIs.
Because the density of noncondensed magnons is still much larger than that of the magnon condensates in realistic experiments,\cite{demokritov} 
we consider two kinds of hybrid FI junctions. The first one (Fig. \ref{fig:system}) is a junction between quasi-equilibrium magnon condensate  and magnons being pumped by microwaves, while the second one (Fig. \ref{fig:System_NC_P}) is a junction between  pumped magnons and noncondensed magnons.
In order to lighten notations, we denote them as the C-P (Fig. \ref{fig:system}) and NC-P (Fig. \ref{fig:System_NC_P}) junctions, with  `C' for quasi-equilibrium magnon condensates, `P' for pumped magnons, and `NC' for noncondensed magnons.


This paper is organized as follows. In Sec. \ref{sec:sn}, we first focus on a single FI (P) and microscopically revisit microwave pumping\cite{mod2,battery} at finite (room) temperature from the viewpoint of  magnon injection\cite{ultrahot}. 
In Sec. \ref{sec:NC/P}, we  apply these ideas to  C-P and NC-P junctions and show that dc magnon currents can arise from an applied ac magnetic field (i.e., microwaves). The amount is drastically enhanced by the ferromagnetic resonance (FMR).
We also address the distinctions between the C-P, NC-P, and C-C junctions.
Lastly in Sec. \ref{sec:persistent}, using a magnon current through the A-C phase, we point out the possibility that such realized magnon currents flow persistently even at finite temperature due to microwave pumping. The predicted amount turns out to be much larger, about $10^3$ times, than the magnon-BEC current predicted in Ref. [\onlinecite{KKPD}].
Finally, Sec. \ref{sec:summary} summarizes our results.

\section{C-P junction}
\label{sec:sn}

We consider the C-P junction shown in Fig. \ref{fig:system} between two FIs, the left FI (C) being in a quasi-equilibrium magnon-BEC\cite{demokritov,ultrahot}, while magnons are being pumped into the right FI (P) by the applied microwave field\cite{ISHE1}. We assume that the magnons in the left FI have already reached a quasi-equilibrium condensate state (i.e., after microwave pumping) through a procedure such as realized in Ref. [\onlinecite{demokritov}], while we continue to apply a microwave only to the right FI and generate continuously magnons.

\subsection{Spin Hamiltonian}
\label{subsec:Hamiltonian}
Each three-dimensional FI depicted in Fig. \ref{fig:system} is described by a Heisenberg spin Hamiltonian   in the presence of a Zeeman term
(we assume a cubic lattice),
\begin{eqnarray}
    {\cal{H}}_{\rm{FI}} =  - \sum_{\langle i j\rangle} {\bf S}_i \cdot {\bf J} \cdot {\bf S}_j -  g\mu_{\rm{B}}  {\bf B}_{{\rm{L(R)}}} \cdot \sum_i {\bf S}_i,
\label{eqn:Heisenberg}
\end{eqnarray}
where ${\bf J}$ denotes a diagonal $3\times 3$-matrix with $\textrm{diag}({\bf J}) = J\{ 1, 1,\eta\}$. 
The exchange interaction between neighboring spins in the ferromagnetic insulator is $J >0$,
$  \eta $ denotes the anisotropy of the spin Hamiltonian,
and ${\bf B}_{\rm{L(R)}} = B_{\rm{L(R)}} {\bf e}_z$ is an applied magnetic field to the left (right)  FI (${\bf e}_z$ denotes the unit vector along the $z$-axis).
The symbol $\langle i j\rangle$ indicates summation over neighboring spins in each FI,
 ${\bf S}_i$ denotes the spin of length $S$ at lattice site $i$,
 and we assume large \cite{uchidainsulator,adachi}  spins $S\gg 1$.
We then use the Holstein-Primakoff  transformation,\cite{HP} $ S_i^+ = \sqrt{2S}[1-a_i^\dagger a_i / (2S)]^{1/2} a_i $  and  $S_i^z = S - a_i^\dagger a_i$, and map Eq. (\ref{eqn:Heisenberg}) onto a system of magnons that satisfy $[a_{i}, a_{j}^{\dagger }] = \delta_{i,j}$.
Using the long wave-length approximation and taking the continuum limit, the energy dispersion relation of magnons in the isotropic system (i.e. $\eta = 1$) becomes 
$ \omega _{{{\bf {k}}}} = 2J S \alpha^2k^2 +g\mu_{\rm{B}}B_{\rm{L(R)}}\equiv D k^2 +g\mu_{\rm{B}}B_{\rm{L(R)}} $, 
where  $\alpha $ denotes the lattice constant, $\mid {\bf {k}} \mid  \equiv k$, and $D\equiv 2J S \alpha^2$. 

Within our microscopic calculation (see Appendix for details), we find that in the continuum limit, the magnon-magnon interaction $  {\cal{H}}_{\rm{FI}}^{\rm{m{\mathchar`-}m}}$ could arise from the $\eta \not=1$ anisotropic spin Hamiltonian
 ${\cal{H}}_{\rm{FI}} $ as the ${\cal{O}}(1-\eta)$ term,  
\begin{eqnarray}
  {\cal{H}}_{\rm{FI}}^{\rm{m{\mathchar`-}m}}  
=  - J_{\rm{m{\mathchar`-}m}} \alpha^3  \int  d{\mathbf{r}} \   a^\dagger (\mathbf{r})  a^\dagger (\mathbf{r})  a (\mathbf{r})  a(\mathbf{r}),
\label{eqn:MagnonMagnonint}
\end{eqnarray}
where $ J_{\rm{m{\mathchar`-}m}} \equiv  - J (1-\eta ) = {\cal{O}}(S^0)$ and 
$ [a(\mathbf{r}), a^\dagger ({\mathbf{r}}^{\prime})]=  \delta ({\mathbf{r}} - {\mathbf{r}}^{\prime})$.
Therefore the magnon-magnon interaction does not influence the magnon transport between $ \eta =1 $ isotropic FIs in any significant manner\cite{KKPD,Kevin2} 
and we can neglect them in the isotropic case (see also Sec. \ref{subsubsec:MagMag}). 
Due to the next leading $1/S$-expansion of the Holstein-Primakoff transformation,
magnon-magnon interactions of the type $a^\dagger a^\dagger a a ={\cal{O}}(S^{0})$
arise also from the hopping terms as well as from the potential term (i.e., $z$-component).
Thus the magnon-magnon interaction [Eq. (\ref{eqn:MagnonMagnonint})] is given by 
$ J_{\rm{m{\mathchar`-}m}} =  - J (1-\eta ) $ in the continuum limit (see Appendix \ref{sec:anisotropyMM} for the details).
We mention that the microscopic relation between the spin Hamiltonian ${\cal{H}}_{\rm{FI}}$ [Eq. (\ref{eqn:Heisenberg})] and the Gross-Pitaevskii (GP) Hamiltonian that describes the Josephson effect in magnon-BECs has already been addressed in Ref. [\onlinecite{KKPD}] (see also the Appendix) and we add it here too for readability.
The GP Hamiltonian actually shows that the magnon-magnon interactions do not influence transport of condensed magnons in such isotropic C-P junctions.

Due to a finite overlap of the wave functions,  there exists in general a finite exchange interaction between the spins located at the boundaries of the different FIs.
Thus, only the boundary spins, denoted as  ${\bf S}_{\Gamma_{\rm L}}$ and ${\bf S}_{\Gamma_{\rm{R}}}$ in the left and right FI, respectively (see Fig. \ref{fig:system}), are relevant to transport of magnons in the junction. The exchange interaction  between the two FIs may thus be described\cite{KKPD}  by the Hamiltonian 
 \begin{eqnarray}
{\cal{H}}_{\rm{ex}} = -J_{\rm{ex}} \sum_{\langle \Gamma_{\rm{L}} \Gamma_{\rm{R}} \rangle} {\bf S}_{\Gamma_{\rm{L}}} \cdot {\bf S}_{\Gamma_{\rm{R}}}, 
\label{eqn:Heisenberg2}
\end{eqnarray}
where $ J_{\rm{ex}} > 0$ is the magnitude of the  exchange interaction at the interface. The two FIs are assumed to be weakly exchange-coupled, {\it i.e.}, $J_{\rm{ex}}\ll J$.
In terms of magnon operators, ${\cal{H}}_{\rm{ex}}$ can be rewritten as 
\begin{equation}
 {\cal{H}}_{\rm{ex}} = -J_{\rm{ex}} S  \sum_{\langle {\Gamma_{\rm L}} {\Gamma_{\rm R}} \rangle}  (a_{\Gamma_{\rm L}} a_{\Gamma_{\rm R}}^{\dagger } + a_{\Gamma_{\rm L}}^{\dagger } a_{\Gamma_{\rm R}})+ {\cal{O}}(S^{0}).
 \label{eqn:exchange}
\end{equation}
We have ignored terms arising from the $z$-component of the spin variables in the Hamiltonian ${\cal{H}}_{\rm{ex}}$, since these do not influence the dynamics of the junction.\cite{KKPD} Finally, the total Hamiltonian ${\cal{H}} $ that describes transport of magnons in the junction reads $ {\cal{H}} =  {\cal{H}}_{\rm{FI}}  + {\cal{H}}_{\rm{ex}} $.  

Using the Heisenberg equation of motion, the time-evolution under ${\cal{H}} $ of the magnon density operator $I$ in the left FI  becomes 
\begin{eqnarray}
I \equiv  \dot{n}_{\rm{L}}(t)  = -i {J_{\rm{ex}}  S}  a_{{\rm{L}}}(t) a_{{\rm{R}}}^{\dagger }(t)/\hbar  + {\rm{H.c.}},
 \label{eqn:current}
\end{eqnarray}
where  $ n_{\rm{L}}(t) \equiv  a_{{\rm{L}}}^{\dagger }(t)a_{{\rm{L}}}(t)$ is the operator of the magnon number density. 
The operator $I$ characterizes the exchange of spin angular momentum via magnons per unit time between the FIs. Therefore (in that sense) it could be regarded as the magnon current density operator in the FI junction.

\subsection{Microwave pumping}
\label{subsec:pumping}

To clarify the features of transport of magnons in the C-P junction, we first focus on the right FI (the pumped magnon part of the C-P junction) and microscopically revisit microwave pumping\cite{mod2,battery,ISHE1} from the viewpoint of magnon injection\cite{demokritov,ultrahot} (Fig. \ref{fig:BEC}).
The interaction between the spins in the FI (P) and the applied microwave (i.e. clockwise rotating magnetic field) may be described by the Hamiltonian\cite{takayoshi}
\begin{eqnarray}\label{eq:vac}
V_{\rm{ac}} = - g\mu_{\rm{B}} \Gamma _0  \sum_j [{\rm{e}}^{i(\Omega t +\vartheta_{\rm{ac}})} S^{+}_j +{\rm{e}}^{-i(\Omega t+\vartheta_{\rm{ac}})} S^{-}_j], 
\label{eqn:ac_microwave}
\end{eqnarray}
where $ \Gamma _0 (\ll  B_{\rm{L(R)}})$ is the magnitude of the applied magnetic microwave field,  $ \Omega $ represents the frequency, and $\vartheta_{\rm{ac}}$ the phase of the microwave field that depends on the initial condition. The summation is over all spins in the right FI (P). Using the Holstein-Primakoff transformation, we rewrite Eq. (\ref{eq:vac})  in terms of magnon operators,~\footnote{Since we apply a weak microwave $\mid  \Gamma _0/B_{\rm{L(R)}}  \mid \sim 10^{-5}$, the Holstein-Primakoff transformation is applicable.} 
\begin{eqnarray}
V_{\rm{ac}} &=& - \sqrt{2S} g\mu_{\rm{B}} \Gamma _0 \sum_j [a_j{\rm{e}}^{i(\Omega t+\vartheta_{\rm{ac}})}  +a_j^{\dagger }{\rm{e}}^{-i(\Omega t+\vartheta_{\rm{ac}})}]    \nonumber   \\
&+& {\cal{O}}(S^{-1/2}). 
\label{eqn:ac_microwave_magnon}
\end{eqnarray}

\begin{figure}[h]
\begin{center}
\includegraphics[width=7.5cm,clip]{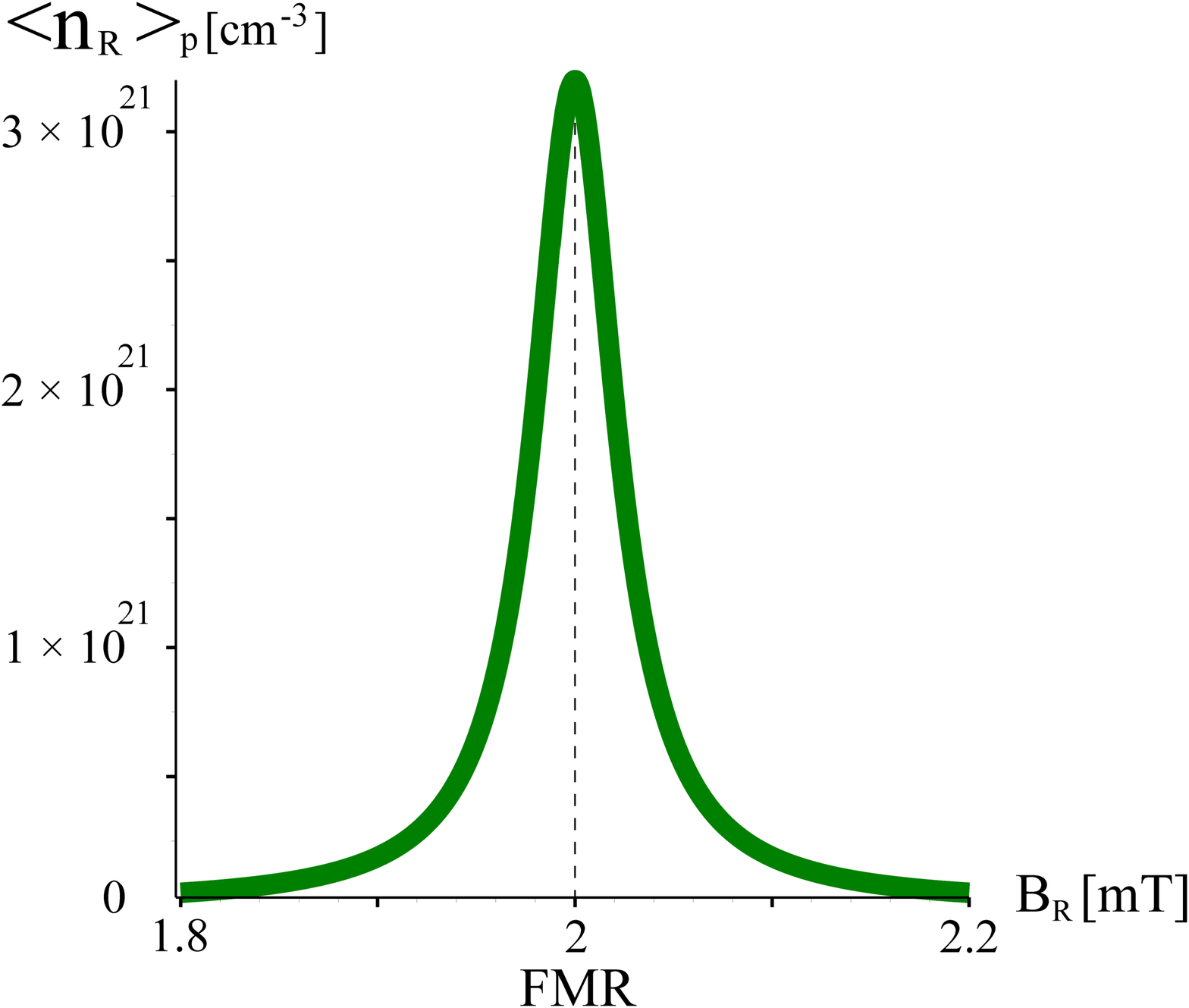}
\caption{(Color online)
A plot of the number density of pumped magnons $ \langle n_{\rm{R}} \rangle_{\rm{P}}$ [cm$^{-3}$] as function of the magnetic field $B_{\rm{R}}$ [mT] obtained by numerically solving Eq. (\ref{eqn:FMR_number}).
As an example, for $ \Omega =400 $MHz, $\Gamma _0=0.25\mu $T, $ \tau_{{\rm{p}}} =100$ns, $\alpha  \approx 1$\AA, $g=2$, and $S=16$,
the number density of pumped magnons amounts to $ \langle n_{\rm{R}} \rangle_{\rm{P}}=3.2 \times  10^{21} $ cm$^{-3}$ on FMR [$\hbar \Omega = g\mu_{\rm{B}}B_{\rm{R}}$ (i.e., $B_{\rm{R}}=2$mT)].
\label{fig:FMR} }
\end{center}
\end{figure}

\subsubsection{FMR}
\label{subsubsec:FMR}

Using again the Heisenberg equation of motion, the time-evolution of the magnon density $ n_{\rm{R}}(t) \equiv  a_{{\rm{R}}}^{\dagger }(t)a_{{\rm{R}}}(t)$ under microwave pumping [i.e. $V_{\rm{ac}}$ in Eq. (\ref{eqn:ac_microwave_magnon})] becomes 
\begin{eqnarray}
 \langle  \dot{n}_{\rm{R}}(t) \rangle_{\rm{P}} = - i  \frac{\sqrt{2\tilde{S} }  g\mu_{\rm{B}} \Gamma _0}{\hbar}\langle a_{\rm{R}}(t)\rangle _{\rm{P}} \ {\rm{e}}^{i(\Omega t+\vartheta_{\rm{ac}})} + {\rm{c.c.}},
 \label{eqn:pumping_rate}
  \end{eqnarray}
where $\tilde{S} \equiv S/\alpha^3  $. 
The subscript `P' in $ \langle a_{\rm{R}}(t)\rangle _{\rm{P}}$ means the expectation value under microwave pumping $V_{\rm{ac}}$.
Treating $V_{\rm{ac}}$ as the perturbative term,
the standard procedure\cite{QSP} of the Schwinger-Keldysh formalism\cite{rammer,tatara} gives
\begin{eqnarray}
\langle  a_{\rm{R}} (t) \rangle_{\rm{P}} &=&  i    \sqrt{2 \tilde{S} }g\mu_{\rm{B}} \Gamma _0 \int  d{\mathbf{r}}^{\prime}
                                                 \int_{\rm{c}} d \tau^{\prime}   \nonumber  \\
                                       &\times &    \langle {\rm{T_c}} a_{\rm{R}} (\tau)  a_{\rm{R}}^{\dagger } (\tau^{\prime})     \rangle  
                                                    {\rm{e}}^{-i(\Omega \tau^{\prime}+\vartheta_{\rm{ac}})}/\hbar   
                                 +  {\cal{O}}  \big(( \Gamma _0)^3 \big),  \nonumber
\end{eqnarray}
where $\tau$ and $\tau^{\prime} $ denote the contour variables defined on the  Schwinger-Keldysh closed time path\cite{rammer,tatara} c, and $\text{T}_{\text{c}}$ is the path-ordering operator defined on it.
Using the Langreth\cite{rammer,tatara} method, it becomes
\begin{eqnarray}
\langle  a_{\rm{R}} (t) \rangle_{\rm{P}}  &=&   -  \sqrt{2 \tilde{S} }g\mu_{\rm{B}} \Gamma _0 \int   d{\mathbf{r}}^{\prime}
                                         \int dt^{\prime}     \nonumber  \\
                            &\times &    [{\cal{G}}_{\rm{R}}^{\rm{t}}(t, t^{\prime}) -  {\cal{G}}_{\rm{R}}^<(t, t^{\prime})]    
                                              {\rm{e}}^{-i(\Omega t^{\prime}+\vartheta_{\rm{ac}})}/\hbar , 
\end{eqnarray}
where  ${\cal{G}}^{\rm{t(<, r)}}({\mathbf{r}}, t, {\mathbf{r}}^{\prime}, {t}^{\prime}) \equiv {\cal{G}}^{\rm{t(<, r)}}(t, {t}^{\prime}) $
 is the bosonic time-ordered (lesser, retarded) Green function, which satisfies  $ {\cal{G}}^{\rm{r}}= {\cal{G}}^{\rm{t}} -  {\cal{G}}^<  $.
After Fourier transformation, it becomes
\begin{eqnarray}
\langle  a_{\rm{R}} (t) \rangle_{\rm{P}} = - \sqrt{2 \tilde{S} }g\mu_{\rm{B}} \Gamma _0    {\rm{e}}^{ - i (\Omega t+\vartheta_{\rm{ac}})}  
                                               {\cal{G}}_{{\rm{R}}, k=0, \Omega}^{\rm{r}}.
\label{eqn:zeropumping}
\end{eqnarray}
Phenomenologically introducing\cite{tatara} a finite life-time\cite{demokritov} $ \tau_{{\rm{p}}} $ ($\sim  100 $ns)  of the pumped magnons mainly due to nonmagnetic impurity scatterings, it becomes  
\begin{equation}
\langle  a_{\rm{R}} (t) \rangle_{\rm{P}} =  \frac{- \sqrt{2 \tilde{S} }g\mu_{\rm{B}} \Gamma _0 }
{ \hbar  \Omega -g\mu_{\rm{B}}B_{\rm{R}} + \frac{i \hbar}{2\tau_{{\rm{p}}}} }  {\rm{e}}^{ - i (\Omega t+\vartheta_{\rm{ac}})}.
 \label{eqn:a_pumping}
\end{equation}
The zero-mode population of magnons is generated by microwave pumping [Eq. (\ref{eqn:zeropumping})] and the number density of the pumped magnons $\langle n_{\rm{R}} \rangle_{\rm{P}} = \langle  a_{\rm{R}} (t) \rangle_{\rm{P}}^{\ast }  \langle a_{\rm{R}} (t) \rangle_{\rm{P}} $ reads 
\begin{eqnarray}
\langle n_{\rm{R}} \rangle_{\rm{P}} =
\frac{2 \tilde{S} (g\mu_{\rm{B}} \Gamma _0)^2 } { (\hbar  \Omega -g\mu_{\rm{B}}B_{\rm{R}})^2 + \Big(\frac{ \hbar}{2\tau_{{\rm{p}}}}\Big)^2}.
 \label{eqn:FMR_number}
 \end{eqnarray}
In terms of spin variables, Eq. (\ref{eqn:a_pumping}) shows a homogeneous macroscopic coherent precession\cite{bunkov} with the frequency $\Omega $ which can be semiclassically\cite{KKPD} treated. The applied microwave forces spins to precess coherently at the same frequency $\Omega $ with the microwave. Thus, we can control the frequency of the homogeneous coherent precession through the applied microwave field.

Eq. (\ref{eqn:FMR_number})  shows that the number density of pumped magnons strongly increases at the FMR\cite{ISHE1} (Fig. \ref{fig:FMR}) 
 \begin{eqnarray}
\hbar \Omega = g\mu_{\rm{B}}B_{\rm{R}}.
 \label{eqn:FMR}
 \end{eqnarray}
 With the help of Eqs. (\ref{eqn:pumping_rate}) and (\ref{eqn:a_pumping}), the resulting magnon pumping rate becomes 
\begin{equation}
\langle  \dot{n}_{{\rm{R}}}  \rangle _{\rm{P}} =  8\tau_{{\rm{p}}} \tilde{S} ({g\mu_{\rm{B}}\Gamma _0}/{\hbar })^2.
 \label{eqn:rate_value}
\end{equation}
Assuming\cite{demokritov} $\Gamma _0=0.25\mu $T, $ \tau_{{\rm{p}}} =100$ns, $\alpha  \approx 1$\AA, $g=2$, and\cite{adachi,uchidainsulator} $S=16$, the magnon pumping rate amounts to   $\langle  \dot{n}_{{\rm{R}}}  \rangle _{\rm{P}} \sim  10^{28} $ cm$^{-3}$s$^{-1}$ at FMR.
Thus, using microwave pumping, we can continuously inject zero-mode magnons into the right FI (P).

\subsubsection{Magnon-magnon interaction}
\label{subsubsec:MagMag}

We next consider the finite (room) temperature effect on microwave pumping.
We remark that a theory\cite{battery,mod2} on microwave pumping at zero-temperature has been considered before, but to the best of our knowledge not yet for finite temperatures.

As remarked in Sec. \ref{subsec:Hamiltonian} (see Appendix \ref{sec:anisotropyMM} for the details of the derivation), 
our microscopic calculation actually shows that in the continuum limit, the magnon-magnon interaction $  {\cal{H}}_{\rm{FI}}^{\rm{m{\mathchar`-}m}}$ could arise from the $\eta \not=1 $ anisotropic spin Hamiltonian ${\cal{H}}_{\rm{FI}} $ as the ${\cal{O}}(1-\eta)$ term 
[see Eqs. (\ref{eqn:Heisenberg}) and (\ref{eqn:MagnonMagnonint})];  
 ${\cal{H}}_{\rm{FI}}^{\rm{m{\mathchar`-}m}}  =  
- J_{\rm{m{\mathchar`-}m}} \alpha^3  \int  d{\mathbf{r}} \   a^\dagger (\mathbf{r})  a^\dagger (\mathbf{r})  a (\mathbf{r})  a(\mathbf{r})$
with $ J_{\rm{m{\mathchar`-}m}} \equiv  - J (1-\eta) = {\cal{O}}(S^0)$ and 
$ [a(\mathbf{r}), a^\dagger ({\mathbf{r}}^{\prime})]=  \delta ({\mathbf{r}} - {\mathbf{r}}^{\prime})$.
Assuming an anisotropic\cite{KKPD,Kevin2} exchange interaction $\eta \not =1$ among the nearest neighboring spins, 
in addition to $V_{\rm{ac}} ={\cal{O}}(S^{1/2})$ [Eq. (\ref{eqn:ac_microwave_magnon})], 
magnon-magnon interactions $ {\cal{H}}_{\rm{FI}}^{\rm{m{\mathchar`-}m}}= {\cal{O}}(S^0)$ thus arise\cite{KKPD,Kevin2}. 
The magnitude  of the magnon-magnon interaction 
$ \mid J_{\rm{m{\mathchar`-}m}} \mid  =  \mid  J (1-\eta) \mid $ depends on the anisotropy of the exchange interaction in the right FI, and we assume 
\begin{equation}\label{eqn:anisotropy}
\mid  J_{\rm{m{\mathchar`-}m}}   \mid   \    \ll   J,  
\end{equation}
to be treated perturbatively (i.e., a small anisotropy enough to become $\mid  \eta -1   \mid  \ll 1$). 
Taking this magnon-magnon interaction into account, we evaluate the expectation value $\langle  a_{\rm{R}} (t) \rangle_{\rm{P}}^{\rm{m{\mathchar`-}m}}$ and clarify the finite temperature effect on microwave pumping.

Following the same procedure with Sec. \ref{subsubsec:FMR},
a straightforward calculation\cite{QSP} based on Schwinger-Keldysh formalism\cite{rammer,tatara} gives 
\begin{eqnarray}
\langle  a_{\rm{R}} (t) \rangle_{\rm{P}}^{\rm{m{\mathchar`-}m}}   
&=&
\frac{4 i  \sqrt{2 \tilde{S} }  J_{\rm{m{\mathchar`-}m}} g\mu_{\rm{B}} \Gamma _0 \alpha ^3}{\hbar ^2}  
\int  d{\mathbf{r}}^{\prime} \int  d{\mathbf{r}}^{\prime  \prime}
\nonumber   \\
&\times & 
 \int  d{t}^{\prime} \int  d{t}^{\prime  \prime}  
{\rm{e}}^{ - i (\Omega t^{\prime }+\vartheta_{\rm{ac}})}
{\cal{G}}^{\rm{<}}_{\rm{R}} ({t}^{\prime  \prime}, {t}^{\prime  \prime}) 
\nonumber   \\
&\times & 
\Big[
{\cal{G}}_{\rm{R}}^{\rm{t}}(t, {t}^{\prime  \prime}){\cal{G}}_{\rm{R}}^{\rm{t}}({t}^{\prime  \prime}, {t}^{\prime })
-{\cal{G}}_{\rm{R}}^{\rm{<}}(t, {t}^{\prime  \prime}){\cal{G}}_{\rm{R}}^{\rm{>}}({t}^{\prime  \prime}, {t}^{\prime })   \nonumber   \\
&-&
{\cal{G}}_{\rm{R}}^{\rm{t}}(t, {t}^{\prime  \prime}){\cal{G}}_{\rm{R}}^{\rm{<}}({t}^{\prime  \prime}, {t}^{\prime })
+{\cal{G}}_{\rm{R}}^{\rm{<}}(t, {t}^{\prime  \prime}){\cal{G}}_{\rm{R}}^{\rm{\bar{t} }}({t}^{\prime  \prime}, {t}^{\prime })
\Big]
 \nonumber    \\
 &+& {\cal{O}}(J_{\rm{m{\mathchar`-}m}}^2),
  \nonumber 
 \label{eqn:a_pumpingMagMag_calculation}
\end{eqnarray}
where  ${\cal{G}}({\mathbf{r}}, t, {\mathbf{r}}^{\prime}, {t}^{\prime  \prime}) \equiv {\cal{G}}(t, {t}^{\prime  \prime}) $ and ${\cal{G}}^{\rm{t(\bar{t}, <, >)}} $ is the bosonic time-ordered (anti-time ordered, lesser, greater) Green function. Using the retarded (advanced) Green function $ {\cal{G}}^{\rm{r(a)}}$, they satisfy\cite{rammer,tatara} $ {\cal{G}}^{\rm{t}} = {\cal{G}}^{\rm{r}} + {\cal{G}}^{\rm{<}}  $, $ {\cal{G}}^{\rm{\bar{t} }} = {\cal{G}}^{\rm{<}} - {\cal{G}}^{\rm{a}}  $, and ${\cal{G}}^{\rm{<}} - {\cal{G}}^{\rm{>}} = {\cal{G}}^{\rm{a}} - {\cal{G}}^{\rm{r}}  $. Consequently, after the Fourier transformation, it becomes
\begin{eqnarray}
\langle  a_{\rm{R}} (t) \rangle_{\rm{P}}^{\rm{m{\mathchar`-}m}}   
&=&
4 \sqrt{2 \tilde{S} }  J_{\rm{m{\mathchar`-}m}} g\mu_{\rm{B}} \Gamma _0 \alpha ^3 
\big[\sum_{\mathbf{k}} f_{\rm{B}}(\omega_{\mathbf{k}}^{\rm{R}})/V\big]
 \nonumber  \\
 &\times & \big({\cal{G}}_{{\rm{R}}, k=0, \Omega }^{\rm{r}}\big)^2 {\rm{e}}^{ - i (\Omega t+\vartheta_{\rm{ac}})},
 \label{eqn:a_pumpingMagMag_calculation5}
\end{eqnarray}
where $ f_{\rm{B}}(\omega_{\mathbf{k}}^{\rm{R}})=({\rm{e}}^{\beta \omega_{\mathbf{k}}^{\rm{R}}}-1)^{-1}$ is the Bose-distribution function, $\beta \equiv 1/(k_{\rm{B}} T)$ the inverse temperature, and $V$ the volume of the system (here the right FI).
We now assume room temperature $ T\simeq 300 $K, which implies $ f_{\rm{B}}(\omega_{\mathbf{k}}^{\rm{R}})\simeq  k_{\rm{B}} T/\omega_{\mathbf{k}}^{\rm{R}}$. Using this approximation, Eq. (\ref{eqn:a_pumpingMagMag_calculation5}) becomes
\begin{equation}
\langle  a_{\rm{R}} (t) \rangle_{\rm{P}}^{\rm{m{\mathchar`-}m}} \approx  \frac{4 \sqrt{2 \tilde{S} }  J_{\rm{m{\mathchar`-}m}} g\mu_{\rm{B}} \Gamma _0     \gamma \alpha^3   k_{\rm{B}} T}
{ \Big(\hbar  \Omega -g\mu_{\rm{B}}B_{\rm{R}} + \frac{i \hbar}{2\tau_{{\rm{p}}}} \Big)^2}  {\rm{e}}^{ - i (\Omega t+\vartheta_{\rm{ac}})},
 \label{eqn:a_pumpingMagMag}
\end{equation}
where the constant $\gamma$  is defined by $  \sum_{\mathbf{k}}   f_{\rm{B}}(\omega_{\mathbf{k}}^{\rm{R}} )/V \equiv   \gamma  k_{\rm{B}} T $. 
Assuming  $ J = 0.1$eV, $B_{\rm{R}} = 50 $mT, $\alpha  \simeq  1$\AA, $g=2$, and\cite{adachi,uchidainsulator}  $S=15$, it amounts to $ \gamma \alpha^3  \simeq  5 \times  10^{-4} $eV$^{-1}$.

Eq. (\ref{eqn:a_pumpingMagMag}) shows that even in the presence of the magnon-magnon interaction $J_{\rm{m{\mathchar`-}m}} $, the spin dynamics under microwave pumping essentially remains the same as the previous one [Eq. (\ref{eqn:a_pumping})] (i.e., without magnon-magnon interactions); localized spins precess coherently with the frequency $\Omega $. 
The significant distinction, however, is that in addition to the zero-mode of magnons ${\cal{G}}_{{\rm{R}}, k=0, \Omega }^{\rm{r}} $ that arises from the applied microwave, the non-zero modes of magnons $  \sum_{\mathbf{k}}   f_{\rm{B}}(\omega_{\mathbf{k}}^{\rm{R}} ) $ are excited by the magnon-magnon interaction $J_{\rm{m{\mathchar`-}m}} $ [Eq. (\ref{eqn:a_pumpingMagMag_calculation5})]. 
The magnon-magnon interaction thus characterizes the finite temperature effect on microwave pumping,
\begin{equation}
\langle  a_{\rm{R}} (t) \rangle_{\rm{P}}^{\rm{m{\mathchar`-}m}} \propto T.
\end{equation}
That is, due to the magnon-magnon interaction $J_{\rm{m{\mathchar`-}m}} $, the non-zero modes of magnons are excited and consequently, the quantity $\langle  a_{\rm{R}} (t) \rangle_{\rm{P}}^{\rm{m{\mathchar`-}m}}$ becomes proportional to the temperature $T$ (for high temperatures).
The number density of such pumped magnons reads 
\begin{equation}
\langle n_{\rm{R}} \rangle_{\rm{P}}^{\rm{m{\mathchar`-}m}}=\mid  \langle a_{\rm{R}} (t) \rangle_{\rm{P}}^{\rm{m{\mathchar`-}m}}  \mid^2 \   \propto T^2.
 \label{eqn:MagMagTemp}
\end{equation}
This means that the number density of pumped magnons due to the magnon-magnon interaction increases with temperature as $ T^2$ in the high temperature regime considered here.
We remark that since 
\begin{equation}
\langle n_{\rm{R}} \rangle_{\rm{P}}^{\rm{m{\mathchar`-}m}} ={\cal{O}}({ J_{\rm{m{\mathchar`-}m}}}^2), 
\end{equation}
the temperature dependence in Eq. (\ref{eqn:MagMagTemp}) is independent of the sign of  $J_{\rm{m{\mathchar`-}m}} $ in Eq. (\ref{eqn:MagnonMagnonint}) and thus holds for repulsive and attractive magnon-magnon
interactions alike.

In conclusion, in the presence of the magnon-magnon interaction $ J_{\rm{m{\mathchar`-}m}}$ [Eq. (\ref{eqn:MagnonMagnonint})], the total number density of pumped magnons at high temperatures becomes [Eqs. (\ref{eqn:FMR_number}) and (\ref{eqn:MagMagTemp})] 
\begin{equation}
\langle n_{\rm{R}} \rangle_{\rm{P}} + \langle n_{\rm{R}} \rangle_{\rm{P}}^{\rm{m{\mathchar`-}m}},  
\nonumber
\end{equation}
where $\langle n_{\rm{R}} \rangle_{\rm{P}}={\cal{O}}(J_{\rm{m{\mathchar`-}m}}^0 T^0)$ and $\langle n_{\rm{R}} \rangle_{\rm{P}}^{\rm{m{\mathchar`-}m}}={\cal{O}}(J_{\rm{m{\mathchar`-}m}}^2 T^2)$. The amount of such pumped magnons becomes maximally at the FMR ($\hbar \Omega = g\mu_{\rm{B}}B_{\rm{R}} $) and it increases with increasing temperature .
Assuming\cite{demokritov,adachi,uchidainsulator} $\alpha =1 $\AA, $S=16 $, $g  =2  $, $\tau_{{\rm{p}}} = 100 $ns, $\Gamma _0 = 0.5 \mu $T, $ \gamma \alpha^3  \simeq  5 \times  10^{-4} $ev$^{-1}$, $T= 300$K, and $J_{\rm{m{\mathchar`-}m}}=25 \mu $eV, 
we get from Eqs. (\ref{eqn:FMR_number}) and (\ref{eqn:MagMagTemp})  $ \langle n_{\rm{R}} \rangle_{\rm{P}} \sim  1 \times  10^{22}$ cm$^{-3}$ and $\langle n_{\rm{R}} \rangle_{\rm{P}}^{\rm{m{\mathchar`-}m}}  \sim   2 \times  10^{21}$ cm$^{-3}$ at the FMR. 

\subsection{Nonequilibrium magnon currents}
\label{subsec:magnon-BEC}

Next, we apply the results obtained in Sec. \ref{subsec:pumping} to the hybrid  C-P junction (Fig. \ref{fig:system}); the junction between the magnon-BEC in the left FI (C) and the pumped magnons  in the right FI (P). 
Assuming again  isotropic\cite{KKPD,Kevin2} exchange interaction  $ \eta =1 $ in ${\cal{H}}_{\rm{FI}}$ [Eq. (\ref{eqn:Heisenberg})] among the nearest neighboring spins whose spin length are large enough\cite{adachi,uchidainsulator}  [i.e. $S \gg 1$], we continue to apply microwave only to the right FI (P) and clarify the features of magnon transport\cite{MagnonSupercurrent,ThermalizationHill}.
We remark that in terms of spin variables, both correspond to a macroscopic coherent precession.\cite{bunkov} A distinct difference, however, is that the frequency of quasi-equilibrium magnon-BECs is essentially characterized\cite{KKPD} by the applied magnetic field $B_{\rm{L}} $ in the left FI (C), while that of pumped magnons by the frequency $\Omega $ of the applied microwave [Eq. (\ref{eqn:a_pumping})]. This plays the key role in the ac-dc conversion of the magnon current we describe below. The number of magnons in the C-P junction is not conserved due to the magnon injection through microwave pumping. Therefore the Josephson equation of the C-C junction microscopically derived in Ref. [\onlinecite{KKPD}] is not directly applicable to the C-P junction.

Since both the quasi-equilibrium magnon condensate and the pumped magnons correspond to  macroscopic coherent precession\cite{bunkov} in terms of spin variables, they may be treated semiclassically\cite{KKPD}. Therefore, following Ref.~[\onlinecite{bender}] the interaction between them may be described by the Hamiltonian [Eq. (\ref{eqn:exchange})],
\begin{eqnarray}
{\cal{H}}_{\rm{ex}}^{\rm{C{\mathchar`-}P}} =  -J_{\rm{ex}} S  \sum_{\langle {\Gamma_{\rm L}} {\Gamma_{\rm R}} \rangle}  
 [\langle a_{\Gamma_{\rm L}} \rangle_{\rm{C}}  \langle a_{\Gamma_{\rm R}}^{\dagger }\rangle_{\rm{P}}  + {\rm{c.c.}}],
  \label{eqn:C/P_Hamiltonian}
 \end{eqnarray}
where\cite{KKPD} $\langle a_{\Gamma_{\rm L}} \rangle_{\rm{C}}$ is the expectation value in the quasi-equilibrium magnon-BEC state.
Under the Hamiltonian ${\cal{H}}_{\rm{ex}}^{\rm{C{\mathchar`-}P}}$, a nonequilibrium magnon current arises from microwave pumping in the C-P junction [Eq. (\ref{eqn:current})],
\begin{eqnarray}
    \langle I (t) \rangle_{\rm{C{\mathchar`-}P}}  =  -i {J_{\rm{ex}}  S  } \langle a_{{\rm{L}}}(t)\rangle_{\rm{C}} \langle a_{{\rm{R}}}^{\dagger }(t)\rangle_{\rm{P}}/{\hbar}+ {\rm{c.c.}},
     \label{eqn:C/Pcurrent}
\end{eqnarray}
where $\langle a_{{\rm{L}}}(t) \rangle_{\rm{C}} = \sqrt{n_{{\rm{L}}}(t)} {\rm{exp}}[-i \vartheta_{{\rm{L}}}(t)]  $. The magnon number density in the left FI is $n_{{\rm{L}}}(t)$ and the phase $\vartheta_{{\rm{L}}} (t)$. 
They are characterized by the GP Hamiltonian that can be microscopically derived from the spin Hamiltonian ${\cal{H}}_{\rm{FI}}$ (see Eq. (\ref{eqn:Heisenberg}) in Ref. [\onlinecite{KKPD}]). The GP Hamiltonian actually shows that {\it only} the homogeneous condensates  play a role, without space dependence of $\langle a_{{\rm{L}}}(t) \rangle_{\rm{C}}$. 
Associated with the macroscopic coherence, the magnon current $\langle I(t) \rangle_{\rm{C{\mathchar`-}P}}$ arises from the ${\cal{O}}(J_{\rm{ex}})$-term. This is the same with the Josephson magnon current\cite{KKPD} in BECs addressed in Ref. [\onlinecite{KKPD}].

We further remark that there could be some contributions in Eq. (\ref{eqn:C/Pcurrent}) [or Eq. (\ref{eqn:exchange})] that might arise from higher order terms in the Holstein-Primakoff expansion\cite{HP} [e.g. $ {\cal{O}}(S^{0})$ and $ {\cal{O}}(S^{-1})$]. Such contributions can be taken into account by using the semiclassical relation $ \langle S_i^+\rangle_{\rm{C(P)}} = \sqrt{2S}[1-\langle a_i^\dagger \rangle_{\rm{C(P)}}  \langle a_i \rangle_{\rm{C(P)}} / (2S)]^{1/2} \langle  a_i \rangle_{\rm{C(P)}}$. 
We have  actually confirmed [Eq. (\ref{eqn:a_pumpingMagMag})] that the main mechanism responsible for the magnon currents remains in place even in the presence of such higher order terms. In addition, we now assume a large spin\cite{demokritov,adachi,uchidainsulator} $S \sim  10$. Therefore, this is enough to focus on the magnon current $\langle I \rangle_{\rm{C{\mathchar`-}P}}={\cal{O}}(S)$ expressed by Eq. (\ref{eqn:C/Pcurrent}).

\begin{figure}[h]
\begin{center}
\includegraphics[width=8.5cm,clip]{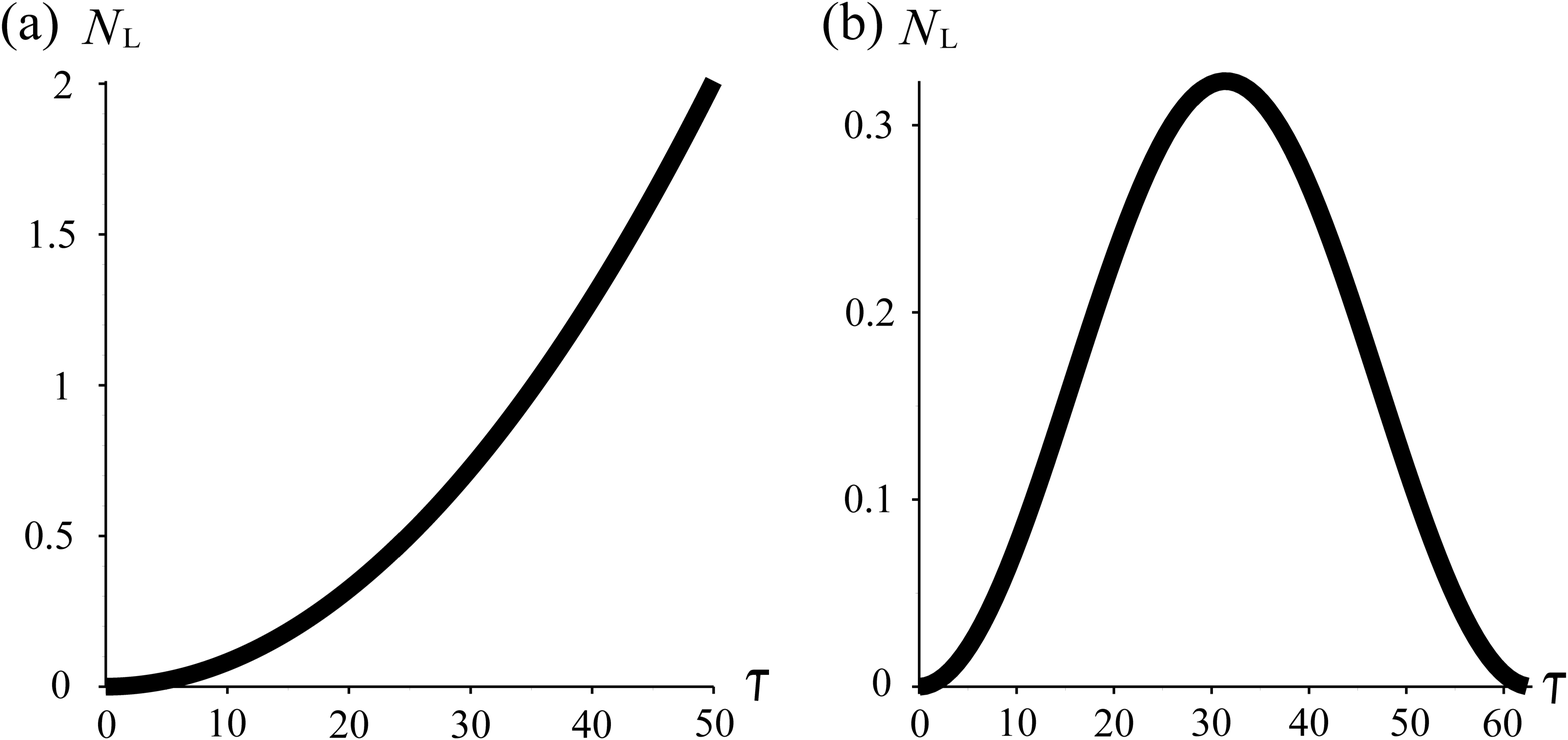}
\caption{
Plots of the rescaled magnon number $N_{\rm{L}}\equiv n_{\rm{L}}  {\alpha^3 } $ as function of the rescaled time $ {\cal{T}}  \equiv  (J_{\rm{ex}}S /\hbar)t  $ obtained by numerically solving Eq. (\ref{eqn:ac/dcJosephson-like}) for the values $g=2$, $S=16$, $J_{\rm{ex}}\approx  0.1 \mu  $eV, $\Gamma _0=0.125\mu $T, $ \tau_{{\rm{p}}} =100$ns, $\alpha  \approx 1 $\AA, $ \vartheta_{{\rm{L}}}(0) -\vartheta_{\rm{ac}} = \pi$, $ N_{\rm{L}} ({\cal{T}}=0) =  10^{-5} $, and $ \Omega =400 $MHz. The rescaled time $ {\cal{T}}=1$ corresponds to $t = 5 $ns and $N_{\rm{L}} =1$  to $n_{\rm{L}} =10^{24}$cm$^{-3} $. $dN_{\rm{L}}/(d{\cal{T}}) $ represents the magnon current.
(a) $ \hbar \Omega = g\mu_{\rm{B}} B_{{\rm{L}}}$ (i.e. $B_{{\rm{L}}} =2$mT). Since the sign of the function $dN_{\rm{L}}/(d{\cal{T}}) $ does not change all the time, it can be regarded as a dc magnon current.
(b)  $ \hbar \Omega \not= g\mu_{\rm{B}} B_{{\rm{L}}}$: $B_{{\rm{L}}} =2.05$mT. This gives the ac magnon current whose period becomes of the order of $10^{-1}\mu $s. 
\label{fig:acdc} }
\end{center}
\end{figure}     

Using Eqs. (\ref{eqn:a_pumping}) and (\ref{eqn:FMR}), the term $\langle  a_{\rm{R}} (t) \rangle_{\rm{P}}$ in Eq. (\ref{eqn:C/Pcurrent}) on FMR reads
\begin{equation}
\langle  a_{\rm{R}} (t) \rangle_{\rm{P}} =  \frac{2i \tau_{{\rm{p}}}}{\hbar }  \sqrt{2 \tilde{S} }g\mu_{\rm{B}} \Gamma _0 
  {\rm{e}}^{ - i (\Omega t+\vartheta_{\rm{ac}})}.
 \label{eqn:a_pumping_FMR}
\end{equation}
Consequently, the magnon current density [Eq. (\ref{eqn:C/Pcurrent})] becomes 
 \begin{eqnarray}
\langle I \rangle_{\rm{C{\mathchar`-}P}} &=& - \frac{4\tau_{{\rm{p}}} J_{\rm{ex}}  S
                   \sqrt{2\tilde{S} n_{{\rm{L}}} }g\mu_{\rm{B}} \Gamma_0 }{\hbar^2}     \nonumber \\
                 & \times  & {\rm{cos}} [ \vartheta_{{\rm{L}}}(t) - \Omega t - \vartheta_{\rm{ac}}].
                 \label{eqn:C/Pcurrent2}
 \end{eqnarray}   
 This is the general expression of the magnon current density in the C-P junction.

We recall that the frequency of the coherent precession due to the magnon-BEC $\vartheta_{{\rm{L}}}(t)$ can be essentially characterized by the effective magnetic field along the $z$-axis,\cite{KKPD} and the magnitude can be controlled by the applied magnetic field, 
 \begin{eqnarray}
 \dot{ \vartheta}_{{\rm{L}}} (t) =  g \mu_{\rm{B}}  B_{{\rm{L}}}/\hbar. 
 \label{eqn:frequency_BEC}
 \end{eqnarray}     
Thus, by tuning it to the microwave frequency (Fig. \ref{fig:system})
 \begin{eqnarray}
  g \mu_{\rm{B}}  B_{{\rm{L}}}/\hbar= \Omega, 
  \label{eqn:condition_dc}
 \end{eqnarray}
the magnon current density becomes
 \begin{eqnarray}
 \langle I \rangle _{\rm{C{\mathchar`-}P}}   &=& - \frac{4\tau_{{\rm{p}}} J_{\rm{ex}}  S}{\hbar^2 }
                                                         \sqrt{2\tilde{S} n_{{\rm{L}}} }g\mu_{\rm{B}} \Gamma_0   \nonumber    \\
&\times & {\rm{cos}} [ \vartheta_{{\rm{L}}}(0) - \vartheta_{\rm{ac}}].
                    \label{eqn:dc-ac_MagnonPumping}
\end{eqnarray}
This result proves that by adjusting the frequency of the applied microwave in the right FI to that of the magnon-BEC in the left FI,
a dc magnon current can arise from an applied ac magnetic field [Fig. \ref{fig:acdc} (a)]. 

We note that the dc magnon current arises not from the FMR condition given in Eq. (\ref{eqn:FMR}) (see Fig. \ref{fig:FMR}), $ g \mu_{\rm{B}}  B_{{\rm{R}}}= \hbar \Omega$,
but from the condition between magnon-BECs and microwaves [Eq. (\ref{eqn:condition_dc}) and Fig. \ref{fig:acdc} (a)], $ g \mu_{\rm{B}}  B_{{\rm{L}}}= \hbar \Omega$.
That is, if only the frequency of the applied microwave in the right FI  is adjusted to that of the  magnon-BEC in the left FI  [Eq. (\ref{eqn:condition_dc})], a dc magnon current can arise. The condition for FMR [Eq. (\ref{eqn:FMR})] is not responsible for the ac-dc conversion of the magnon current
but it does lead to its 
drastic enhancement, see Fig. \ref{fig:FMR}.

Assuming\cite{demokritov,uchidainsulator,adachi}  $J_{\rm{ex}}\approx  0.1 \mu  $eV,  $\Gamma _0=0.125\mu $T, $ \tau_{{\rm{p}}} =100$ns, $\alpha  \approx 1 $\AA, $g=2$, $S=16$, $ \vartheta_{{\rm{L}}}(0) -\vartheta_{\rm{ac}} \approx 0$, and\footnote{Since Serga {\textit{et al.}}\cite{ultrahot} have reported that the decay time of quasi-equilibrium magnon condensate is much longer than that of pumped magnons, we here assume that the life-time of quasi-equilibrium magnon condensate  is infinite.} $ n_{\rm{L}}\sim  10^{19}  \  {\rm{cm}}^{-3}$, the magnitude of the dc magnon current [Eq. (\ref{eqn:dc-ac_MagnonPumping})] becomes $\mid \langle I \rangle_{\rm{C{\mathchar`-}P}} \mid  \sim   10^{29}   {\rm{cm}}^{-3}s^{-1}$. Due to the magnon injection through microwave pumping and the resulting FMR, it is worth stressing that this value is actually much larger, about $10^{6}$ times, than the one obtained from the dc Josephson magnon current in the C-C junction (the `C-C' junction denotes the FI junction between magnon condensates (C) that was introduced in Ref. [\onlinecite{KKPD}]).  

The magnon currents can be experimentally measured by using the same method\cite{magnon2,dipole} as for the Josephson magnon current described in Ref.  [\onlinecite{KKPD}]. Since the moving magnetic dipoles (i.e., magnons with  magnetic moment $g\mu_{\rm{B}}  {\bf e}_z$) produce electric fields, magnon currents can be observed by measuring the resulting voltage drop\footnote{Regarding the detail, see Ref.  [\onlinecite{KKPD}].}, which is proportional to the amount of the magnon current. 
This means that the voltage drop in the C-P junction  becomes much larger, about $10^{6}$ times, than the one in the C-C junction.\cite{KKPD} Therefore we expect that the measurement of the magnon currents in the C-P junction is experimentally
better accessible than the Josephson magnon currents in the C-C junction.  We remark that applying an electric field to the interface and using the resulting A-C  phase as in Ref. [\onlinecite{KKPD}], we could tune the magnitude of the phase in the cosine function of Eq. (\ref{eqn:dc-ac_MagnonPumping}) and consequently, we might control the direction of the dc magnon current.

\subsection{Distinction from C-C junction}
\label{subsec:Distinction}

Introducing the rescaled time $ {\cal{T}} $  and the number of magnons $N_{\rm{L}}$ by  $ t \equiv   \hbar {\cal{T}} /(J_{\rm{ex}}S) $ and $ n_{\rm{L}} \equiv  N_{\rm{L}}/{\alpha^3 }$, Eq. (\ref{eqn:C/Pcurrent2}) can be rewritten as [with the help of Eqs. (\ref{eqn:frequency_BEC}) and (\ref{eqn:current})]
 \begin{eqnarray}
 \frac{d N_{\rm{L}}({\cal{T}})}{d {\cal{T}} }  &=& - \frac{4\sqrt{2S}\tau_{{\rm{p}}} g\mu_{\rm{B}} \Gamma_0}{\hbar}
                                                         \sqrt{N_{{\rm{L}}}({\cal{T}}) }   \label{eqn:ac/dcJosephson-like}     \\
&\times & {\rm{cos}} \Big[\frac{g \mu_{\rm{B}}  B_{{\rm{L}}} -\hbar  \Omega}{J_{\rm{ex}}S}  {\cal{T}}  
+ \vartheta_{{\rm{L}}}(0) - \vartheta_{\rm{ac}}\Big]. \nonumber 
\end{eqnarray}
Fig. \ref{fig:acdc} shows the results obtained by numerically solving Eq. (\ref{eqn:ac/dcJosephson-like}).
Based on them, we summarize the qualitative distinction between the magnon currents in the C-P junction and those in the C-C junction\cite{KKPD}.

\subsubsection{dc magnon currents}
\label{subsubsec:dc}

The frequency of the macroscopic coherent spin precession of magnon-BECs is characterized by the magnetic field along the $z$-axis. Through a Berry phase\cite{Mignani,Hea} referred to as A-C phase,\cite{casher} the dc Josephson magnon current in the C-C junction arises\cite{KKPD} from an applied time-dependent magnetic field (for details, see Ref. [\onlinecite{KKPD}]). In that case, the magnetic field gradient has to be proportional to the time,
 \begin{eqnarray}
 B_{{\rm{L}}} -   B_{{\rm{R}}} \propto  t    \   \
   ({\rm{i.e.,}}  \   B_{{\rm{L}}} \not=  B_{{\rm{R}}}).
   \label{eqn:dc_C/C}
 \end{eqnarray}
 This is the condition for the dc Josephson magnon current in the C-C junction.

On the other hand, the C-P junction does not require such conditions. The applied microwave forces spins to precess coherently [Eq. (\ref{eqn:a_pumping})]. The frequency of the macroscopic coherent precession in the FI (P) is characterized not by the magnetic field $g\mu_{\rm{B}}B_{{\rm{R}}}/\hbar $, but by the applied microwave $\Omega $. Therefore by tuning $\Omega = g\mu_{\rm{B}}B_{{\rm{L}}}/\hbar$,
a dc magnon current can arise [Fig. \ref{fig:acdc} (a)]. The relation between each magnetic field $B_{{\rm{L}}}  $ and $B_{{\rm{R}}}  $ has nothing to do with the dc magnon current, which is in sharp contrast to the C-C junction [Eq. (\ref{eqn:dc_C/C})].

We note that by adjusting the magnetic field of the right FI (P), $g\mu_{\rm{B}}B_{{\rm{R}}}/\hbar =  \Omega  \  (=g\mu_{\rm{B}}B_{{\rm{L}}}/\hbar )$, FMR does occur and consequently, the pumped magnon drastically increases (Fig. \ref{fig:FMR}). This leads to a drastic enhancement of the dc magnon current. 
In conclusion, through FMR, a dc magnon current of the C-P junction can arise from the condition [Fig. \ref{fig:acdc} (a)],
$  B_{{\rm{L}}} = B_{{\rm{R}}}  =   \hbar \Omega/(g\mu_{\rm{B}})$,
which is in sharp contrast to the C-C junction\cite{KKPD} [Eq. (\ref{eqn:dc_C/C})].

\subsubsection{ac magnon currents}
\label{subsubsec:ac}

The above discussion is applicable also to the distinction between the ac magnon current in the C-C junction\cite{KKPD} and that in the C-P junction.
In the C-C junction, an ac Josephson magnon current arises when a static magnetic field gradient is produced ($ {\dot{B} }_{{\rm{L(R)}}}=0$) $ B_{{\rm{L}}} -   B_{{\rm{R}}}= ({\rm{constant}}) \not=0 $. On the other hand, this condition does not necessarily manifest the ac magnon current in the C-P junction. Even in that case, a dc magnon current arises (due to the same reason explained above) if $B_{{\rm{L}}}  =  \hbar \Omega/(g\mu_{\rm{B}}) $. 

Fig. \ref{fig:acdc} (b) is an example of the ac magnon current in the C-P junction obtained by numerically solving Eq. (\ref{eqn:ac/dcJosephson-like}). Assuming $g=2$, $ \Omega =400 $MHz, and $B_{{\rm{L}}}=2.05$mT [i.e. $B_{{\rm{L}}}  \not=  \hbar \Omega/(g\mu_{\rm{B}}) $], the period of the oscillation becomes of the order of $10^{-1}\mu $s, which is within experimental reach. The period can be controlled by tuning $\Omega $ or $ g\mu_{\rm{B}}B_{{\rm{L}}}/\hbar $.

\begin{figure}[h]
\begin{center}
\includegraphics[width=8cm,clip]{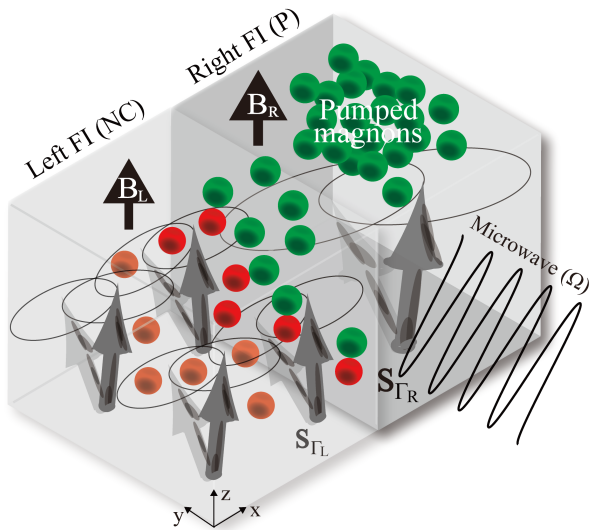}
\caption{(Color online)
Schematic representation of the NC-P junction, consisting of a left FI with noncondensed magnons (NC) and a weakly exchange coupled right FI with pumped magnons (P).
The microwave at frequency $\Omega $ is applied only to the right FI. The distinction to the C-P junction in Fig. \ref{fig:system} is that the magnons in the left FI are not condensed (i.e.,  $\langle  a_{\rm{L}} \rangle =0 $). Due to the exchange interaction ${\cal{H}}_{\rm{ex}}$ [Eq. ({\ref{eqn:exchange}})], the applied microwave to the right FI can indirectly affect the noncondensed magnons of the left FI and thereby generate an indirect FMR there.
\label{fig:System_NC_P} }
\end{center}
\end{figure}

\subsection{Elastic scatterings}
\label{subsec:elastic}

In this paper we consider insulators having the parabolic dispersion relation [i.e., $\omega _{{{\bf {k}}}} = D k^2 +g\mu_{\rm{B}}B_{\rm{L(R)}}$] where the zero-mode $k=0$ is the lowest energy state and therefore we have assumed that magnon-BECs are formed in the zero-mode \cite{bunkov}.
We have then seen that the ac and dc magnon transport properties are characterized by the macroscopic coherence of pumped and condensed magnons in the zero-mode $\langle a_0  \rangle \not=0$.
In such C-P junctions, those zero-mode magnons are exchanged in ${\cal{O}}(J_{\rm{ex}})$, and actually only the kinetic momentum conserved scatterings at the interface play an essential role; 
for elastic scattering, the momentum non-conserved scatterings indeed cannot contribute to  transport due to the macroscopic coherence.
Therefore in this paper we have focused on the momentum conserved scatterings.
Despite the enormous experimental challenge, we believe that such magnon-BECs in the zero-mode would be experimentally realizable.

We mention, however, that  taking into account inelastic scatterings, 
our description of magnon transport ${\cal{O}}(J_{\rm{ex}})$ might in principle be applicable also to the general BECs in the non-zero modes ($k=k_{\rm{min}} \not=0$). 
The frequency of the $k=0$ BECs we have used is [Eq. (\ref{eqn:frequency_BEC})] $g\mu_{\rm{B}} B_{\rm{L}}/\hbar$ where $B_{\rm{L}}$ is the external magnetic field, while that of the general $k=k_{\rm{min}}$ BECs becomes \cite{bunkov} $g\mu_{\rm{B}} B_{\rm{eff}}/\hbar$ where $g\mu_{\rm{B}} B_{\rm{eff}}= g\mu_{\rm{B}} B_{\rm{L}} + Dk_{\rm{min}}^2$.
Taking into account inelastic scatterings and replacing $B_{\rm{L}}$ by $B_{\rm{eff}}$,
our description for $O(J_{\rm{ex}})$ magnon transport would be applicable also to the general BECs.

\section{NC-P junction}
\label{sec:NC/P}

In the last section (Sec. \ref{sec:sn}), we have seen that quasi-equilibrium magnon condensate generates ac and dc  magnon currents from the applied ac magnetic field (i.e., microwave). Microwave pumping and the resulting FMR (Fig. \ref{fig:FMR}) drastically enhance the amount of such magnon currents.

In this section, we further develop it by taking  experimental results into account that\cite{demokritov} the density of noncondensed magnons is still much larger than that of the magnon condensate, and thus propose an alternative  way to generate a dc magnon current and enhance it. 

To this end, we replace the magnon-BECs of the C-P junction (Fig. \ref{fig:system}) with noncondensed\cite{bunkov} magnons $\langle  a_{\rm{L}} \rangle =0 $ as shown in Fig. \ref{fig:System_NC_P}. Thus we newly consider the hybrid ferromagnetic insulating junction between noncondensed magnons and pumped magnons, namely, the NC-P junction (Fig. \ref{fig:System_NC_P}). 
The only difference to the C-P junction (Sec. \ref{sec:sn}) is that the magnons in the left FI are not condensed $\langle  a_{\rm{L}} \rangle =0 $.

Then, using the microscopic description of microwave pumping addressed in Sec. \ref{subsec:pumping}, 
we clarify the features of the magnon current in the  $\eta =1 $ isotropic NC-P junction (Fig. \ref{fig:System_NC_P}). Focusing on the effect of the noncondensed magnons on the ac/dc conversion of the magnon current, we then discuss the difference from the C-P and C-C junctions\cite{KKPD}.

\subsection{dc magnon current}
\label{subsec:MagnonCurrent_NC/P}

Following the same procedure with Sec. \ref{subsec:pumping}, a straight-forward calculation based on the Schwinger-Keldysh\cite{rammer,tatara} formalism gives the magnon current $\langle I \rangle _{{\rm{NC{\mathchar`-}P}}} $ on FMR ($\hbar \Omega  = g\mu_{\rm{B}}  B_{\rm{R}}$) in the NC/P junction 
 \begin{eqnarray}
\langle I \rangle _{{\rm{NC{\mathchar`-}P}}}
&=& 2 {\rm{Re}} \Big[ \frac{i}{\hbar } \big( 2 J_{\rm{ex}}  S \tau_{\rm{p}}  \sqrt{2\tilde{S} } g\mu_{\rm{B}} \Gamma _0 \big)^2
            {\cal{G}}^{\rm{r}}_{{\rm{L}}, k=0, \Omega } \Big]   \nonumber     \\
&+& {\cal{O}}(J_{\rm{ex}}^4)  +  {\cal{O}}(J_{\rm{ex}}^2S^2), 
\label{eqn:magnoncurrent_NCP}
\end{eqnarray}
where ${\cal{G}}^{\rm{r}}$ is the bosonic retarded Green function and $\langle I \rangle _{{\rm{NC{\mathchar`-}P}}}= {\cal{O}}(J_{\rm{ex}}^2S^3)$.
Since we assume a weak exchange coupling  $J_{\rm{ex}}/J \sim  10^{-6}$ and $\eta =1$ isotropic large\cite{demokritov,uchidainsulator,adachi} spins $S \gg  1$,
any higher perturbation terms in terms of $J_{\rm{ex}}$ and those due to the $1/S$-expansion of the Holstein-Primakoff transformation
[i.e., $ {\cal{O}}(J_{\rm{ex}}^4)$ and ${\cal{O}}(J_{\rm{ex}}^2S^2)$]
do not influence in any significant manner.
We have actually confirmed that compared with that $\langle I \rangle _{{\rm{NC{\mathchar`-}P}}}$ [Eq. (\ref{eqn:magnoncurrent_NCP})],
the ${\cal{O}}(J_{\rm{ex}}^2S^2)$ term arising from the next leading $1/S$-expansion
becomes quite small, about  $ 10^{-6}$ times, and it is indeed negligible.

We then phenomenologically introduce\cite{tatara} a life-time $ \tau_{\rm{NC}} $  of noncondensed magnons in the left FI mainly due to nonmagnetic impurity scatterings into the Green function ${\cal{G}}^{\rm{r}}$ (in the same way with Sec. \ref{subsec:pumping}). Consequently, the magnon current density becomes
 \begin{eqnarray}
\langle I \rangle_{{\rm{NC{\mathchar`-}P}}}
= \frac{1}{\hbar^2 } \frac{\tau_{\rm{p}}^2}{\tau_{\rm{NC}}}
\frac{\big(2 \sqrt{2\tilde{S} }   J_{\rm{ex}}  S g\mu_{\rm{B}} \Gamma _0\big)^2}{(\hbar \Omega  - g\mu_{\rm{B}} B_{\rm{L}})^2 +  [\hbar /(2 \tau_{\rm{NC}})]^2}.
\label{eqn:magnoncurrent_NCP2}
 \end{eqnarray}
 Due to the effect that the magnons in the left FI is not condensed $\langle  a_{\rm{L}} \rangle =0 $, the magnon current arises from the ${\cal{O}}(J_{\rm{ex}}^2)$-term and it becomes essentially the dc magnon current. These are in sharp contrast\cite{KKPD} to the C-P and C-C junctions.

Tuning
  \begin{eqnarray}
\hbar \Omega  = g\mu_{\rm{B}}  B_{\rm{L}},
\label{eqn:IndirectFMR}
 \end{eqnarray}
the amount $\langle I \rangle_{{\rm{NC{\mathchar`-}P}}}  $ becomes maximally.

\subsection{Indirect FMR}
\label{subsec:IndirectFMR}
 
Eq. (\ref{eqn:magnoncurrent_NCP2}) indicates that tuning $\hbar \Omega  = g\mu_{\rm{B}}  B_{\rm{L}}$, an {\textit{indirect}} FMR does occur in the left FI. Any microwaves are not directly applied to the left FI (Fig. \ref{fig:System_NC_P}). Nevertheless, a FMR does occur in the left FI since the applied microwave to the right FI can still affect the spins of the left FI via the exchange interaction ${\cal{H}}_{\rm{ex}}$ [Eq. ({\ref{eqn:exchange}})]. The applied microwave forces the spins of the right FI to precess coherently with the frequency $\Omega $ and excite the zero-mode of magnons [Eqs. (\ref{eqn:pumping_rate}) and (\ref{eqn:zeropumping}) in Sec. \ref{subsec:pumping}]. Those pumped magnons propagate and interact with the (boundary) spins in the left FI via ${\cal{H}}_{\rm{ex}}$, and produce the zero-mode of magnons with the frequency $\Omega $ [i.e. ${\cal{G}}^{\rm{r}}_{{\rm{L}}, k=0, \Omega }$ in Eq. (\ref{eqn:magnoncurrent_NCP})], which have the same frequency with the microwave directly applied to the right FI. Thus, the applied microwave to the right FI can indirectly affect the left FI via the pumped magnons and generate the indirect FMR.
These double FMRs, the direct FMR in the right FI and the indirect one in the left FI, lead to the drastic enhancement of the resulting magnon current.

\subsection{Distinctions from C-C and C-P junctions}
\label{subsec:Distinction2}

Focusing on the distinction from the C-P and C-C junctions, we summarize the features of magnon currents in the NC-P junction (Fig. \ref{fig:System_NC_P}).

The main point is that in the NC-P junction, the magnon currents arise as $\langle I \rangle _{{\rm{NC{\mathchar`-}P}}} = {\cal{O}}(J_{\rm{ex}}^2)$ 
[Eq. (\ref{eqn:magnoncurrent_NCP2})].
This is in sharp contrast\cite{KKPD} to the C-C and C-P junctions.
Therefore even when we apply an electric field to the interface, the resulting A-C phase cannot play any roles on magnon transport in any significant manners. 
This is in sharp contrast to the C-C and C-P junctions, where the A-C phase can play the key role on the ac/dc conversion of the magnon currents ${\cal{O}}(J_{\rm{ex}})$.

Another distinction is that the magnon current of the NC-P junction becomes essentially the dc one [Eq. (\ref{eqn:magnoncurrent_NCP2})]. 
The noncondensed magnons generate the dc magnon current from the applied ac magnetic field.
Even when a static [i.e., time-independent $d B_{\rm{L(R)}}/(dt) = 0$] magnetic field gradient is  produced, $ B_{\rm{L}} -  B_{\rm{R}} \not= 0$, the resulting magnon current becomes the dc one. This is in sharp contrast to the C-C and C-P junctions, where the magnon current can become the ac one due to the reason addressed in Sec. \ref{subsec:Distinction}. As stressed, the conversion of the ac and dc magnon currents in the C-P junction is characterized not by the magnetic fields, but by the relation between $ g\mu_{\rm{B}}  {B}_{{\rm{L}}}/\hbar $ and $ \Omega $. Therefore, if $ g\mu_{\rm{B}}  {B}_{{\rm{L}}}/\hbar \not= \Omega $, the magnon current becomes an ac one even in that case.

Finally, comparing the maximum magnon current densities available in each hybrid junction,
we clarify the quantitative effect of microwave pumping on the generation of magnon currents;
$ \langle I \rangle _{\rm{NC{\mathchar`-}P}}$, $ \langle I \rangle _{\rm{C{\mathchar`-}P}}$, and the one of the C-C junction\cite{KKPD} $ \langle I \rangle _{\rm{C{\mathchar`-}C}}$.
Assuming\cite{demokritov,uchidainsulator,adachi} $J_{\rm{ex}} = 0.1 \mu  $eV,  $\alpha =1 $\AA, $S=15 $, $g  =2  $, $\tau_{{\rm{p(NC)}}} = 100 $ns, the magnon-BEC density [Eq. (\ref{eqn:C/Pcurrent2})] $ n_{\rm{L}} \sim  10^{19}  \  {\rm{cm}}^{-3}$, and $\Gamma _0 = 0.5 \mu $T, each ratio on the direct and the resulting indirect FMRs [$ B_{{\rm{L}}} = B_{{\rm{R}}}  =   \hbar \Omega/(g\mu_{\rm{B}})$] becomes\cite{KKPD}
\begin{subequations}
 \begin{eqnarray}
 \frac{\mid  \langle I \rangle_{\rm{NC{\mathchar`-}P}}  \mid    }{\mid  \langle I \rangle_{\rm{C{\mathchar`-}P}} \mid    } \simeq  1.1 \times 10^{5},  \\
 \frac{\mid  \langle I \rangle_{\rm{NC{\mathchar`-}P}}  \mid    }{\mid  \langle I \rangle_{\rm{C{\mathchar`-}C}}  \mid    } \simeq  3.6 \times 10^{6}. 
\label{eqn:RateValue}
 \end{eqnarray}
 \end{subequations}
These give
 \begin{eqnarray}
\mid  \langle I \rangle_{\rm{C{\mathchar`-}C}} \mid    \
\ll   \  \mid     \langle I \rangle_{\rm{C{\mathchar`-}P}}  \mid   \
\ll    \   \mid    \langle I \rangle_{\rm{NC{\mathchar`-}P}}  \mid.   
\label{eqn:RateSummary}
 \end{eqnarray} 
This shows that the magnon injection through microwave pumping and the resulting FMRs drastically enhances the magnon current. In addition to the direct FMR, due to the indirect FMR in the adjacent left FI (Sec. \ref{subsec:IndirectFMR}), the magnon current of the NC-P junction (Fig. \ref{fig:System_NC_P}) can become much larger than that of the C-P junction.

\section{Magnon current through Aharonov-Casher phase}
\label{sec:persistent}

Lastly, using microwave pumping, we discuss the possibility that a magnon current through the A-C phase flows persistently\cite{LossPersistent,LossPersistent2,Kopietz} even at finite (room) temperature. 
In Ref. [\onlinecite{KKPD}], using  magnon-BECs\cite{demokritov,ultrahot}, we have found that the A-C phase\cite{casher,Mignani,Hea} generates a persistent magnon-BEC\cite{MagnonSupercurrent} current at zero-temperature. 
Since both magnon condensate and pumped magnons correspond to the macroscopic coherent precession\cite{bunkov} in terms of spin variables, it is expected that a similar\cite{MagnonSupercurrent,ThermalizationHill} magnon current arises from microwave pumping.
Therefore, based on the microscopic description of microwave pumping in Sec. \ref{subsec:pumping}, we consider  finite temperatures and evaluate the magnon current driven by an A-C phase.

To this end, we revisit the discussion of Ref. [\onlinecite{KKPD}], and expand it  by taking the effect of microwave pumping into account. We consider a single FI described by the Hamiltonian $ {\cal{H}}_{\rm{FI}}  $ [Eq. (\ref{eqn:Heisenberg})] and  apply now an electric field $E$ along the $y$-axis $  {\mathbf{E}} ({\bf r})= E{\bf e}_y $ as well as a homogeneous magnetic field $B$ along the $z$-axis. A magnetic dipole $g\mu_{\rm{B}}  {\bf e}_z$ (i.e., magnon) moving along a path $\nu $ in an electric field ${\bf E}({\bf r})$ acquires a Berry phase\cite{Mignani,Hea}, called the A-C phase\cite{casher},
\begin{equation}
 \theta _{\rm{A{\mathchar`-}C}} = \frac{g \mu_{\rm{B}}}{\hbar c^2} \int_{\nu } \textrm{d} {\bf l} \cdot \left[{\bf E}({\bf r}) \times  {\bf e}_z\right]. 
 \label{eqn:A-Cphase}
\end{equation}
Since $  {\mathbf{E}}({\bf r}) = E{\bf e}_y $, it becomes ${\bf E}({\bf r}) \times  {\bf e}_z = E{\bf e}_x$. Therefore we focus on magnon transport along the $x$-axis, which is essentially described\cite{magnon2,KKPD} by the Hamiltonian $ {\cal{H}}_{\rm{FI}}^{\rm{A{\mathchar`-}C}}  
=  - J  S  \sum_{ j }   (a_{j} a_{j + 1}^{\dagger }  {\rm{e}}^{- i  \theta_{\rm{A{\mathchar`-}C}}} + {\rm{H.c.}}) +g\mu_{\rm{B}}B  \sum_{ j }    a_{j}^{\dagger } a_{j}$. The summation is over all spins along the $x$-axis. 
The A-C phase reads $  \theta _{\rm{A{\mathchar`-}C}} = [g \mu_{\rm{B}}/(\hbar c^2)] E \alpha  $.

We then continue to apply microwave $\Omega $ [$V_{\rm{ac}} $ in Eq. (\ref{eqn:ac_microwave_magnon})] and generate a macroscopic coherent precession described by Eq. (\ref{eqn:a_pumping}). 
Consequently, the A-C phase induces a magnon current analogous to the magnon-BEC current\cite{KKPD,MagnonSupercurrent}.
The amount becomes maximally on FMR $\hbar \Omega = g\mu_{\rm{B}}B $ (Fig. \ref{fig:FMR}). 
The resulting magnon current density induced by the A-C phase reads~\cite{KKPD},
\begin{equation}
\langle {\cal{I}}\rangle = \frac{2JS}{\hbar } \langle n \rangle_{\rm{P}}  \  {\rm{sin}} \theta_{\rm{A{\mathchar`-}C}},
 \label{eqn:PumpedMagnonCurrent}
\end{equation}
where $\langle n \rangle_{\rm{P}}  = 8 \tilde{S}  ({\tau_{{\rm{p}}} g\mu_{\rm{B}}\Gamma _0}/{\hbar })^2$ is
the pumped magnon density on FMR [Eq. (\ref{eqn:FMR_number}) and Fig. \ref{fig:FMR}].
In the presence of the magnon-magnon interaction (Sec. \ref{subsubsec:MagMag}), it becomes 
\begin{equation}
\langle {\cal{I}}\rangle = \frac{2JS}{\hbar } 
\Big[\langle n \rangle_{\rm{P}}  + \langle n \rangle_{\rm{P}}^{\rm{m{\mathchar`-}m}} \Big] 
{\rm{sin}} \theta_{\rm{A{\mathchar`-}C}},
 \label{eqn:PumpedMagnonCurrentMagMag}
\end{equation}
where $ \langle n \rangle_{\rm{P}}^{\rm{m{\mathchar`-}m}}  = (16 \sqrt{2 \tilde{S} }  \tau_{{\rm{p}}}^2   J_{\rm{m{\mathchar`-}m}} g\mu_{\rm{B}} \Gamma _0 \gamma \alpha ^3   k_{\rm{B}} T / \hbar ^2)^2  \propto  T^2$ [Eqs. (\ref{eqn:a_pumpingMagMag}) and (\ref{eqn:MagMagTemp})].  

Using the above single FI and microwave pumping, we form a ring consisting of a cylindrical wire analogous to the magnon-BEC ring.
Then, it is expected that the magnon current flows like the persistent\cite{KKPD} magnon-BEC current
(we refer for further details of the magnon-BEC ring where the current is quantized to Ref. [\onlinecite{KKPD}]).
We point out the possibility that due to microwave pumping, the magnon current might flow  even at finite temperatures. 
Dissipations\cite{C-L,CL_text} would arise at finite temperature due to the coupling with thermal bath e.g. due to lattice vibrations (i.e., phonons), but such detrimental effects can be compensated by magnon injection\cite{demokritov,ultrahot} through microwave pumping where the pumping rate [Eqs. (\ref{eqn:pumping_rate}) and (\ref{eqn:rate_value})] is larger than the dissipative decay rate. 

Assuming\cite{demokritov,uchidainsulator,adachi} $\alpha =1 $\AA, $S=16 $, $g  =2  $, $\tau_{{\rm{p}}} = 100 $ns, $\Gamma _0 = 0.5 \mu $T, $ \gamma \alpha^3  \simeq  5 \times  10^{-4} $ev$^{-1}$ (Sec. \ref{subsubsec:MagMag}), $T= 300$K, and $J_{\rm{m{\mathchar`-}m}}=30 \mu $eV, 
they amount to $ \langle n \rangle_{\rm{P}} \simeq  1.28 \times  10^{22}$ cm$^{-3}$ and
$\langle n \rangle_{\rm{P}}^{\rm{m{\mathchar`-}m}}  \simeq   3 \times  10^{21}$ cm$^{-3}$. 
The magnon current [Eq. (\ref{eqn:PumpedMagnonCurrentMagMag})] becomes much larger, about $10^3$ times, than the magnon-BEC current\cite{KKPD}.
Using the same setup (i.e., cylindrical wire) with Ref.  [\onlinecite{KKPD}], the electric field\cite{magnon2,dipole} of the order of mV/m is induced (see Sec. \ref{subsec:magnon-BEC})  and consequently, the voltage drop amounts to of the order of $\mu $V, which is actually much larger than the one arising from the magnon-BEC current ($\sim $nV). Thus, we expect the direct measurement of such magnon currents to be experimentally more realizable.

\section{Summary}
\label{sec:summary}

We have constructed a microscopic theory on magnon transport through microwave pumping in FIs.
Due to the magnon-magnon interaction, the non-zero mode of magnons are excited and the number density of such pumped magnons (P) increases when temperature rises. We have then shown that the magnon injection through such microwave pumping and the resulting ferromagnetic resonance drastically enhances the magnon currents in FIs. 
In hybrid FI junctions, quasi-equilibrium magnon condensates (C) convert the applied ac magnetic field (i.e., microwaves) into an ac and dc magnon current, while noncondensate magnons (NC) convert it into essentially a dc magnon current. 
Due to the direct and indirect FMRs, the NC-P junction becomes the best one (among the C-C, C-P, and NC-P junctions) to generate the largest magnon current.
In a single FI, microwave pumping can produce a magnon current through a Berry phase (i.e., the Aharonov-Casher phase) even at finite temperature in the presence of magnon-magnon interactions. 
The amount of such  magnon currents increases when temperature rises, since the number density of pumped magnons increases.
Due to FMR, the amount becomes much larger, about $10^{3}$ times, than the magnon-BEC current.

\begin{acknowledgments}
We thank K. van Hoogdalem and A. Zyuzin for helpful discussions. 
We acknowledge support by the Swiss NSF, the NCCR QSIT ETHZ-Basel, 
by the JSPS Postdoctoral Fellow for Research Abroad (No. 26-143), Grant-in-Aid for JSPS Research Fellow (No. 25-2747), by the Exchange Programme between Japan and Switzerland 2014 under the Japanese-Swiss Science and Technology Programme supported by JSPS and ETHZ (KN), and
by ANR under Contract No. DYMESYS (ANR 2011-IS04-001-01) (PS). 
\end{acknowledgments}

\appendix*

\section{Spin anisotropy and   \\
magnon-magnon interaction}
\label{sec:anisotropyMM}

In this Appendix, starting from the general Heisenberg spin Hamiltonian, we microscopically provide magnon-magnon interactions in the continuum limit.
To this end, it is enough to focus on the spins aligned along the $x$-axis in each FI and 
the spin Hamiltonian $  {\cal{H}}_{\rm{FI}} $ gives [Eq. (\ref{eqn:Heisenberg})] 
\begin{eqnarray}
    {\cal{H}}_{\rm{FI}}   &=& -J \sum_{ i }  
 {\Big[}   \frac{1}{2} (S_i^{+} S_{i+1}^{-} + S_i^{-} S_{i+1}^{+}) + \eta  S_i^{z} S_{i+1}^{z} {\Big]}   \nonumber    \\
 &-&   g\mu_{\rm{B}} B_{\rm{L(R)}} \sum_{ i } S_i^{z},
\label{eqn:Heisenberg222}
\end{eqnarray}
where the exchange interaction between neighboring spins in the ferromagnetic insulator is $J >0$, $  \eta $ denotes the spin anisotropy,
and  ${\bf B}_{\rm{L(R)}} = B_{\rm{L(R)}} {\bf e}_z$ is the applied magnetic field along the $z$-axis.
The summation is over all spins aligned along the $x$-axis in each FI.
Using the (non-linearized) Holstein-Primakoff  transformation\cite{HP},
\begin{subequations}
\begin{eqnarray}
S_i^+ &= &  \sqrt{2S}\Big(1- \frac{a_i^\dagger a_i}{4 S}\Big) a_i  +{\cal{O}}(S^{-3/2})    \label{eqn:HPnext}    \\
&=& (S_i^-)^\dagger,  \nonumber    \\
S_i^z  &= &  S- a_i^\dagger a_i, 
\end{eqnarray}
\end{subequations}
Eq. (\ref{eqn:Heisenberg222}) provides the magnon-magnon interaction of the form
\begin{eqnarray}
  {\cal{H}}_{\rm{FI}}^{\rm{m{\mathchar`-}m}}  
&=&  \frac{J}{4} \sum_{ i }  
         (a_i^\dagger a_{i+1}^\dagger a_i a_i 
+        a_{i+1}^\dagger a_{i+1}^\dagger a_i a_{i+1}    \nonumber     \\
&+&    a_i^\dagger a_{i+1}^\dagger a_{i+1} a_{i+1}
+        a_i^\dagger a_{i}^\dagger a_i a_{i+1}            \nonumber     \\
&-&   4   \eta  a_i^\dagger a_{i+1}^\dagger a_i a_{i+1}).
\label{eqn:mm2}
\end{eqnarray}
It should be noted that due to the next leading $1/S$-expansion of the Holstein-Primakoff transformation [Eq. (\ref{eqn:HPnext})],
magnon-magnon interactions of the type $a^\dagger a^\dagger a a ={\cal{O}}(S^{0})$
arise also from the hopping terms   (i.e., $S_i^{+} S_{i+1}^{-} +$ H.c.)  of Eq. (\ref{eqn:Heisenberg222})
as well as from the potential term (i.e., $z$-component) $\eta  S_i^{z} S_{i+1}^{z}$.

Finally, taking the continuum limit (i.e. the lattice constant $\alpha \rightarrow 0$) and using the corresponding representation,
$a_i/\sqrt{\alpha } \equiv a(x) - (\alpha /2)(\partial a/\partial x)$ and 
$a_{i+1}/\sqrt{\alpha } \equiv a(x) + (\alpha /2)(\partial a/\partial x)$, 
Eq. (\ref{eqn:mm2}) reduces to
\begin{eqnarray}
  {\cal{H}}_{\rm{FI}}^{\rm{m{\mathchar`-}m}}  
&=&  J (1-\eta )  \alpha  \int  d x  \  a^\dagger (x)  a^\dagger (x)  a (x)  a(x)    \nonumber      \\
&+ &  {\cal{O}}(\alpha ^2),
\label{eqn:MagnonMagnonint000}
\end{eqnarray}
where $ [a(x), a^\dagger (x^{\prime})]=  \delta (x - x^{\prime})$.
Thus in the continuum limit, the magnon-magnon interaction  becomes [Eq. (\ref{eqn:MagnonMagnonint})]
\begin{eqnarray}
  {\cal{H}}_{\rm{FI}}^{\rm{m{\mathchar`-}m}}  
&=&  - J_{\rm{m{\mathchar`-}m}}  \alpha  \int  d x  \  a^\dagger (x)  a^\dagger (x)  a (x)  a(x),
\label{eqn:MagnonMagnonint999}
\end{eqnarray}
with $ J_{\rm{m{\mathchar`-}m}} \equiv  - J (1-\eta ) = {\cal{O}}(S^0)$.
This shows that in the isotropic case $ \eta =1 $, $ J_{\rm{m{\mathchar`-}m}}  = 0$,
and therefore the magnon-magnon interaction does not influence the magnon transport in any significant manner within the continuum limit.

\bibliography{PumpingRef}

\begin{thebibliography}{49}
\expandafter\ifx\csname natexlab\endcsname\relax\def\natexlab#1{#1}\fi
\expandafter\ifx\csname bibnamefont\endcsname\relax
  \def\bibnamefont#1{#1}\fi
\expandafter\ifx\csname bibfnamefont\endcsname\relax
  \def\bibfnamefont#1{#1}\fi
\expandafter\ifx\csname citenamefont\endcsname\relax
  \def\citenamefont#1{#1}\fi
\expandafter\ifx\csname url\endcsname\relax
  \def\url#1{\texttt{#1}}\fi
\expandafter\ifx\csname urlprefix\endcsname\relax\def\urlprefix{URL }\fi
\providecommand{\bibinfo}[2]{#2}
\providecommand{\eprint}[2][]{\url{#2}}

\bibitem[{\citenamefont{Kajiwara et~al.}(2010)\citenamefont{Kajiwara, Harii,
  Takahashi, Ohe, Uchida, Mizuguchi, Umezawa, Kawai, Ando, Takanashi
  et~al.}}]{spinwave}
\bibinfo{author}{\bibfnamefont{Y.}~\bibnamefont{Kajiwara}},
  \bibinfo{author}{\bibfnamefont{K.}~\bibnamefont{Harii}},
  \bibinfo{author}{\bibfnamefont{S.}~\bibnamefont{Takahashi}},
  \bibinfo{author}{\bibfnamefont{J.}~\bibnamefont{Ohe}},
  \bibinfo{author}{\bibfnamefont{K.}~\bibnamefont{Uchida}},
  \bibinfo{author}{\bibfnamefont{M.}~\bibnamefont{Mizuguchi}},
  \bibinfo{author}{\bibfnamefont{H.}~\bibnamefont{Umezawa}},
  \bibinfo{author}{\bibfnamefont{H.}~\bibnamefont{Kawai}},
  \bibinfo{author}{\bibfnamefont{K.}~\bibnamefont{Ando}},
  \bibinfo{author}{\bibfnamefont{K.}~\bibnamefont{Takanashi}},
  \bibnamefont{et~al.}, \bibinfo{journal}{Nature}
  \textbf{\bibinfo{volume}{464}}, \bibinfo{pages}{262} (\bibinfo{year}{2010}).

\bibitem[{\citenamefont{Awschalom and
  Flatt$\acute{\text{e}}$}(2007)}]{awschalom}
\bibinfo{author}{\bibfnamefont{D.~D.} \bibnamefont{Awschalom}}
  \bibnamefont{and} \bibinfo{author}{\bibfnamefont{M.~E.}
  \bibnamefont{Flatt$\acute{\text{e}}$}}, \bibinfo{journal}{Nat. Phys.}
  \textbf{\bibinfo{volume}{3}}, \bibinfo{pages}{153} (\bibinfo{year}{2007}).

\bibitem[{\citenamefont{Zutic et~al.}(2004)\citenamefont{Zutic, Fabian, and
  Sarma}}]{mod}
\bibinfo{author}{\bibfnamefont{I.}~\bibnamefont{Zutic}},
  \bibinfo{author}{\bibfnamefont{J.}~\bibnamefont{Fabian}}, \bibnamefont{and}
  \bibinfo{author}{\bibfnamefont{S.~D.} \bibnamefont{Sarma}},
  \bibinfo{journal}{Rev. Mod. Phys.} \textbf{\bibinfo{volume}{76}},
  \bibinfo{pages}{323} (\bibinfo{year}{2004}).

\bibitem[{\citenamefont{Demokritov et~al.}(2006)\citenamefont{Demokritov,
  Demidov, Dzyapko, Melkov, Serga, Hillebrands, and Slavin}}]{demokritov}
\bibinfo{author}{\bibfnamefont{S.~O.} \bibnamefont{Demokritov}},
  \bibinfo{author}{\bibfnamefont{V.~E.} \bibnamefont{Demidov}},
  \bibinfo{author}{\bibfnamefont{O.}~\bibnamefont{Dzyapko}},
  \bibinfo{author}{\bibfnamefont{G.~A.} \bibnamefont{Melkov}},
  \bibinfo{author}{\bibfnamefont{A.~A.} \bibnamefont{Serga}},
  \bibinfo{author}{\bibfnamefont{B.}~\bibnamefont{Hillebrands}},
  \bibnamefont{and} \bibinfo{author}{\bibfnamefont{A.~N.}
  \bibnamefont{Slavin}}, \bibinfo{journal}{Nature}
  \textbf{\bibinfo{volume}{443}}, \bibinfo{pages}{430} (\bibinfo{year}{2006}).

\bibitem[{\citenamefont{Serga et~al.}(2014)\citenamefont{Serga, Tiberkevich,
  Sandweg, Vasyuchka, Bozhko, Chumak, Neumann, Obry, Melkov, Slavin
  et~al.}}]{ultrahot}
\bibinfo{author}{\bibfnamefont{A.~A.} \bibnamefont{Serga}},
  \bibinfo{author}{\bibfnamefont{V.~S.} \bibnamefont{Tiberkevich}},
  \bibinfo{author}{\bibfnamefont{C.~W.} \bibnamefont{Sandweg}},
  \bibinfo{author}{\bibfnamefont{V.~I.} \bibnamefont{Vasyuchka}},
  \bibinfo{author}{\bibfnamefont{D.~A.} \bibnamefont{Bozhko}},
  \bibinfo{author}{\bibfnamefont{A.~V.} \bibnamefont{Chumak}},
  \bibinfo{author}{\bibfnamefont{T.}~\bibnamefont{Neumann}},
  \bibinfo{author}{\bibfnamefont{B.}~\bibnamefont{Obry}},
  \bibinfo{author}{\bibfnamefont{G.~A.} \bibnamefont{Melkov}},
  \bibinfo{author}{\bibfnamefont{A.~N.} \bibnamefont{Slavin}},
  \bibnamefont{et~al.}, \bibinfo{journal}{Nat. Commun.}
  \textbf{\bibinfo{volume}{5}}, \bibinfo{pages}{3452} (\bibinfo{year}{2014}).

\bibitem[{\citenamefont{Clausen et~al.}({\natexlab{a}})\citenamefont{Clausen,
  Bozhko, Vasyuchka, Melkov, Hillebrands, and Serga}}]{MagnonSupercurrent}
\bibinfo{author}{\bibfnamefont{P.}~\bibnamefont{Clausen}},
  \bibinfo{author}{\bibfnamefont{D.~A.} \bibnamefont{Bozhko}},
  \bibinfo{author}{\bibfnamefont{V.~I.} \bibnamefont{Vasyuchka}},
  \bibinfo{author}{\bibfnamefont{G.~A.} \bibnamefont{Melkov}},
  \bibinfo{author}{\bibfnamefont{B.}~\bibnamefont{Hillebrands}},
  \bibnamefont{and} \bibinfo{author}{\bibfnamefont{A.~A.} \bibnamefont{Serga}},
  \bibinfo{note}{arXiv:1503.00482}.

\bibitem[{\citenamefont{Nakata et~al.}(2014)\citenamefont{Nakata, van
  Hoogdalem, Simon, and Loss}}]{KKPD}
\bibinfo{author}{\bibfnamefont{K.}~\bibnamefont{Nakata}},
  \bibinfo{author}{\bibfnamefont{K.~A.} \bibnamefont{van Hoogdalem}},
  \bibinfo{author}{\bibfnamefont{P.}~\bibnamefont{Simon}}, \bibnamefont{and}
  \bibinfo{author}{\bibfnamefont{D.}~\bibnamefont{Loss}},
  \bibinfo{journal}{Phys. Rev. B} \textbf{\bibinfo{volume}{90}},
  \bibinfo{pages}{144419} (\bibinfo{year}{2014}).

\bibitem[{\citenamefont{Meier and Loss}(2003)}]{magnon2}
\bibinfo{author}{\bibfnamefont{F.}~\bibnamefont{Meier}} \bibnamefont{and}
  \bibinfo{author}{\bibfnamefont{D.}~\bibnamefont{Loss}},
  \bibinfo{journal}{Phys. Rev. Lett.} \textbf{\bibinfo{volume}{90}},
  \bibinfo{pages}{167204} (\bibinfo{year}{2003}).

\bibitem[{\citenamefont{van Hoogdalem and Loss}(2011)}]{Kevin2}
\bibinfo{author}{\bibfnamefont{K.~A.} \bibnamefont{van Hoogdalem}}
  \bibnamefont{and} \bibinfo{author}{\bibfnamefont{D.}~\bibnamefont{Loss}},
  \bibinfo{journal}{Phys. Rev. B} \textbf{\bibinfo{volume}{84}},
  \bibinfo{pages}{024402} (\bibinfo{year}{2011}).

\bibitem[{\citenamefont{Takei and Tserkovnyak}(2014)}]{takei}
\bibinfo{author}{\bibfnamefont{S.}~\bibnamefont{Takei}} \bibnamefont{and}
  \bibinfo{author}{\bibfnamefont{Y.}~\bibnamefont{Tserkovnyak}},
  \bibinfo{journal}{Phys. Rev. Lett.} \textbf{\bibinfo{volume}{112}},
  \bibinfo{pages}{227201} (\bibinfo{year}{2014}).

\bibitem[{\citenamefont{Bender et~al.}(2012)\citenamefont{Bender, Duine, and
  Tserkovnyak}}]{bender}
\bibinfo{author}{\bibfnamefont{S.~A.} \bibnamefont{Bender}},
  \bibinfo{author}{\bibfnamefont{R.~A.} \bibnamefont{Duine}}, \bibnamefont{and}
  \bibinfo{author}{\bibfnamefont{Y.}~\bibnamefont{Tserkovnyak}},
  \bibinfo{journal}{Phys. Rev. Lett.} \textbf{\bibinfo{volume}{108}},
  \bibinfo{pages}{246601} (\bibinfo{year}{2012}).

\bibitem[{\citenamefont{Sch$\ddot{{\text{u}}}$tz
  et~al.}(2003)\citenamefont{Sch$\ddot{{\text{u}}}$tz, Kollar, and
  Kopietz}}]{Kopietz}
\bibinfo{author}{\bibfnamefont{F.}~\bibnamefont{Sch$\ddot{{\text{u}}}$tz}},
  \bibinfo{author}{\bibfnamefont{M.}~\bibnamefont{Kollar}}, \bibnamefont{and}
  \bibinfo{author}{\bibfnamefont{P.}~\bibnamefont{Kopietz}},
  \bibinfo{journal}{Phys. Rev. Lett.} \textbf{\bibinfo{volume}{91}},
  \bibinfo{pages}{017205} (\bibinfo{year}{2003}).

\bibitem[{\citenamefont{Morimae et~al.}(2005)\citenamefont{Morimae, Sugita, and
  Shimizu}}]{morimae}
\bibinfo{author}{\bibfnamefont{T.}~\bibnamefont{Morimae}},
  \bibinfo{author}{\bibfnamefont{A.}~\bibnamefont{Sugita}}, \bibnamefont{and}
  \bibinfo{author}{\bibfnamefont{A.}~\bibnamefont{Shimizu}},
  \bibinfo{journal}{Phys. Rev. A} \textbf{\bibinfo{volume}{71}},
  \bibinfo{pages}{032317} (\bibinfo{year}{2005}).

\bibitem[{\citenamefont{Andrianov and Moiseev}(2014)}]{MagnonQubit}
\bibinfo{author}{\bibfnamefont{S.~N.} \bibnamefont{Andrianov}}
  \bibnamefont{and} \bibinfo{author}{\bibfnamefont{S.~A.}
  \bibnamefont{Moiseev}}, \bibinfo{journal}{Phys. Rev. A}
  \textbf{\bibinfo{volume}{90}}, \bibinfo{pages}{042303}
  (\bibinfo{year}{2014}).

\bibitem[{\citenamefont{Chakraborty et~al.}(2015)\citenamefont{Chakraborty,
  Wenk, and Schliemann}}]{2dMagnonLifetime}
\bibinfo{author}{\bibfnamefont{A.}~\bibnamefont{Chakraborty}},
  \bibinfo{author}{\bibfnamefont{P.}~\bibnamefont{Wenk}}, \bibnamefont{and}
  \bibinfo{author}{\bibfnamefont{J.}~\bibnamefont{Schliemann}},
  \bibinfo{journal}{Eur. Phys. J. B} \textbf{\bibinfo{volume}{88}},
  \bibinfo{pages}{64} (\bibinfo{year}{2015}).

\bibitem[{\citenamefont{Serga et~al.}(2010)\citenamefont{Serga, Chumak, and
  Hillebrands}}]{magnonics}
\bibinfo{author}{\bibfnamefont{A.~A.} \bibnamefont{Serga}},
  \bibinfo{author}{\bibfnamefont{A.~V.} \bibnamefont{Chumak}},
  \bibnamefont{and}
  \bibinfo{author}{\bibfnamefont{B.}~\bibnamefont{Hillebrands}},
  \bibinfo{journal}{J. Phys. D} \textbf{\bibinfo{volume}{43}},
  \bibinfo{pages}{264002} (\bibinfo{year}{2010}).

\bibitem[{\citenamefont{Stamps et~al.}(2014)\citenamefont{Stamps, Breitkreutz,
  Akerman, Chumak, Otani, Bauer, Thiele, Bowen, Majetich, Klaui
  et~al.}}]{spincalreview}
\bibinfo{author}{\bibfnamefont{R.~L.} \bibnamefont{Stamps}},
  \bibinfo{author}{\bibfnamefont{S.}~\bibnamefont{Breitkreutz}},
  \bibinfo{author}{\bibfnamefont{J.}~\bibnamefont{Akerman}},
  \bibinfo{author}{\bibfnamefont{A.~V.} \bibnamefont{Chumak}},
  \bibinfo{author}{\bibfnamefont{Y.}~\bibnamefont{Otani}},
  \bibinfo{author}{\bibfnamefont{G.~E.~W.} \bibnamefont{Bauer}},
  \bibinfo{author}{\bibfnamefont{J.-U.} \bibnamefont{Thiele}},
  \bibinfo{author}{\bibfnamefont{M.}~\bibnamefont{Bowen}},
  \bibinfo{author}{\bibfnamefont{S.~A.} \bibnamefont{Majetich}},
  \bibinfo{author}{\bibfnamefont{M.}~\bibnamefont{Klaui}},
  \bibnamefont{et~al.}, \bibinfo{journal}{J. Phys. D: Appl. Phys.}
  \textbf{\bibinfo{volume}{47}}, \bibinfo{pages}{333001}
  (\bibinfo{year}{2014}).

\bibitem[{\citenamefont{Trauzettel et~al.}(2008)\citenamefont{Trauzettel,
  Simon, and Loss}}]{Trauzettel}
\bibinfo{author}{\bibfnamefont{B.}~\bibnamefont{Trauzettel}},
  \bibinfo{author}{\bibfnamefont{P.}~\bibnamefont{Simon}}, \bibnamefont{and}
  \bibinfo{author}{\bibfnamefont{D.}~\bibnamefont{Loss}},
  \bibinfo{journal}{Phys. Rev. Lett.} \textbf{\bibinfo{volume}{101}},
  \bibinfo{pages}{017202} (\bibinfo{year}{2008}).

\bibitem[{\citenamefont{Clausen et~al.}({\natexlab{b}})\citenamefont{Clausen,
  Bozhko, Vasyuchka, Hillebrands, Melkov, and Serga}}]{ThermalizationHill}
\bibinfo{author}{\bibfnamefont{P.}~\bibnamefont{Clausen}},
  \bibinfo{author}{\bibfnamefont{D.~A.} \bibnamefont{Bozhko}},
  \bibinfo{author}{\bibfnamefont{V.~I.} \bibnamefont{Vasyuchka}},
  \bibinfo{author}{\bibfnamefont{B.}~\bibnamefont{Hillebrands}},
  \bibinfo{author}{\bibfnamefont{G.~A.} \bibnamefont{Melkov}},
  \bibnamefont{and} \bibinfo{author}{\bibfnamefont{A.~A.} \bibnamefont{Serga}},
  \bibinfo{note}{arXiv:1502.07836}.

\bibitem[{\citenamefont{Saitoh et~al.}(2006)\citenamefont{Saitoh, Ueda,
  Miyajima, and Tatara}}]{ISHE1}
\bibinfo{author}{\bibfnamefont{E.}~\bibnamefont{Saitoh}},
  \bibinfo{author}{\bibfnamefont{M.}~\bibnamefont{Ueda}},
  \bibinfo{author}{\bibfnamefont{H.}~\bibnamefont{Miyajima}}, \bibnamefont{and}
  \bibinfo{author}{\bibfnamefont{G.}~\bibnamefont{Tatara}},
  \bibinfo{journal}{Appl. Phys. Lett.} \textbf{\bibinfo{volume}{88}},
  \bibinfo{pages}{182509} (\bibinfo{year}{2006}).

\bibitem[{\citenamefont{Tserkovnyak et~al.}(2005)\citenamefont{Tserkovnyak,
  Brataas, Bauer, and Halperin}}]{mod2}
\bibinfo{author}{\bibfnamefont{Y.}~\bibnamefont{Tserkovnyak}},
  \bibinfo{author}{\bibfnamefont{A.}~\bibnamefont{Brataas}},
  \bibinfo{author}{\bibfnamefont{G.~E.~W.} \bibnamefont{Bauer}},
  \bibnamefont{and} \bibinfo{author}{\bibfnamefont{B.~I.}
  \bibnamefont{Halperin}}, \bibinfo{journal}{Rev. Mod. Phys.}
  \textbf{\bibinfo{volume}{77}}, \bibinfo{pages}{1375} (\bibinfo{year}{2005}).

\bibitem[{\citenamefont{B.-Romanov et~al.}(1984)\citenamefont{B.-Romanov,
  Bunkov, Dmitriev, and Mukharskiy}}]{He3Bunkov}
\bibinfo{author}{\bibfnamefont{A.~S.} \bibnamefont{B.-Romanov}},
  \bibinfo{author}{\bibfnamefont{Y.~M.} \bibnamefont{Bunkov}},
  \bibinfo{author}{\bibfnamefont{V.~V.} \bibnamefont{Dmitriev}},
  \bibnamefont{and} \bibinfo{author}{\bibfnamefont{Y.~M.}
  \bibnamefont{Mukharskiy}}, \bibinfo{journal}{JETP Lett.}
  \textbf{\bibinfo{volume}{40}}, \bibinfo{pages}{1033} (\bibinfo{year}{1984}).

\bibitem[{\citenamefont{Bunkov and Volovik}(2013, arXiv:1003.4889)}]{bunkov}
\bibinfo{author}{\bibfnamefont{Y.~M.} \bibnamefont{Bunkov}} \bibnamefont{and}
  \bibinfo{author}{\bibfnamefont{G.~E.} \bibnamefont{Volovik}},
  \emph{\bibinfo{title}{Novel Superfluids (Chapter IV); eds. K. H. Bennemann
  and J. B. Ketterson}} (\bibinfo{publisher}{Oxford University Press, Oxford},
  \bibinfo{year}{2013, arXiv:1003.4889}).

\bibitem[{\citenamefont{Yukalov}(2012)}]{Yukalov}
\bibinfo{author}{\bibfnamefont{V.~I.} \bibnamefont{Yukalov}},
  \bibinfo{journal}{Laser Phys.} \textbf{\bibinfo{volume}{22}},
  \bibinfo{pages}{1145} (\bibinfo{year}{2012}).

\bibitem[{\citenamefont{Zapf et~al.}(2014)\citenamefont{Zapf, Jaime, and
  Batista}}]{BatistaBEC}
\bibinfo{author}{\bibfnamefont{V.}~\bibnamefont{Zapf}},
  \bibinfo{author}{\bibfnamefont{M.}~\bibnamefont{Jaime}}, \bibnamefont{and}
  \bibinfo{author}{\bibfnamefont{C.~D.} \bibnamefont{Batista}},
  \bibinfo{journal}{Rev. Mod. Phys.} \textbf{\bibinfo{volume}{86}},
  \bibinfo{pages}{563} (\bibinfo{year}{2014}).

\bibitem[{\citenamefont{Demidov et~al.}(2007)\citenamefont{Demidov, Dzyapko,
  Demokritov, Melkov, and Slavin}}]{demidov2}
\bibinfo{author}{\bibfnamefont{V.~E.} \bibnamefont{Demidov}},
  \bibinfo{author}{\bibfnamefont{O.}~\bibnamefont{Dzyapko}},
  \bibinfo{author}{\bibfnamefont{S.~O.} \bibnamefont{Demokritov}},
  \bibinfo{author}{\bibfnamefont{G.~A.} \bibnamefont{Melkov}},
  \bibnamefont{and} \bibinfo{author}{\bibfnamefont{A.~N.}
  \bibnamefont{Slavin}}, \bibinfo{journal}{Phys. Rev. Lett.}
  \textbf{\bibinfo{volume}{99}}, \bibinfo{pages}{037205}
  (\bibinfo{year}{2007}).

\bibitem[{\citenamefont{Vannucchi et~al.}(2010)\citenamefont{Vannucchi,
  $\acute{\rm{A}}$. R.~Vasconcellos, and Luzzi}}]{Vannucchi}
\bibinfo{author}{\bibfnamefont{F.~S.} \bibnamefont{Vannucchi}},
  \bibinfo{author}{\bibnamefont{$\acute{\rm{A}}$. R.~Vasconcellos}},
  \bibnamefont{and} \bibinfo{author}{\bibfnamefont{R.}~\bibnamefont{Luzzi}},
  \bibinfo{journal}{Phys. Rev. B} \textbf{\bibinfo{volume}{82}},
  \bibinfo{pages}{140404(R)} (\bibinfo{year}{2010}).

\bibitem[{\citenamefont{Vannucchi et~al.}(2013)\citenamefont{Vannucchi,
  $\acute{\rm{A}}$. R.~Vasconcellos, and Luzzi}}]{Vannucchi2}
\bibinfo{author}{\bibfnamefont{F.~S.} \bibnamefont{Vannucchi}},
  \bibinfo{author}{\bibnamefont{$\acute{\rm{A}}$. R.~Vasconcellos}},
  \bibnamefont{and} \bibinfo{author}{\bibfnamefont{R.}~\bibnamefont{Luzzi}},
  \bibinfo{journal}{Eur. Phys. J. B} \textbf{\bibinfo{volume}{86}},
  \bibinfo{pages}{463} (\bibinfo{year}{2013}).

\bibitem[{\citenamefont{Aharonov and Casher}(1984)}]{casher}
\bibinfo{author}{\bibfnamefont{Y.}~\bibnamefont{Aharonov}} \bibnamefont{and}
  \bibinfo{author}{\bibfnamefont{A.}~\bibnamefont{Casher}},
  \bibinfo{journal}{Phys. Rev. Lett.} \textbf{\bibinfo{volume}{53}},
  \bibinfo{pages}{319} (\bibinfo{year}{1984}).

\bibitem[{\citenamefont{Mignani}(1991)}]{Mignani}
\bibinfo{author}{\bibfnamefont{R.}~\bibnamefont{Mignani}}, \bibinfo{journal}{J.
  Phys. A: Math. Gen.} \textbf{\bibinfo{volume}{24}}, \bibinfo{pages}{L421}
  (\bibinfo{year}{1991}).

\bibitem[{\citenamefont{Hea and McKellarb}(1991)}]{Hea}
\bibinfo{author}{\bibfnamefont{X.-G.} \bibnamefont{Hea}} \bibnamefont{and}
  \bibinfo{author}{\bibfnamefont{B.}~\bibnamefont{McKellarb}},
  \bibinfo{journal}{Phys. Lett. B} \textbf{\bibinfo{volume}{264}},
  \bibinfo{pages}{129} (\bibinfo{year}{1991}).

\bibitem[{\citenamefont{Loss et~al.}(1990)\citenamefont{Loss, Goldbart, and
  Balatsky}}]{LossPersistent}
\bibinfo{author}{\bibfnamefont{D.}~\bibnamefont{Loss}},
  \bibinfo{author}{\bibfnamefont{P.}~\bibnamefont{Goldbart}}, \bibnamefont{and}
  \bibinfo{author}{\bibfnamefont{A.~V.} \bibnamefont{Balatsky}},
  \bibinfo{journal}{Phys. Rev. Lett.} \textbf{\bibinfo{volume}{65}},
  \bibinfo{pages}{1655} (\bibinfo{year}{1990}).

\bibitem[{\citenamefont{Loss and Goldbart}(1992)}]{LossPersistent2}
\bibinfo{author}{\bibfnamefont{D.}~\bibnamefont{Loss}} \bibnamefont{and}
  \bibinfo{author}{\bibfnamefont{P.~M.} \bibnamefont{Goldbart}},
  \bibinfo{journal}{Phys. Rev. B} \textbf{\bibinfo{volume}{45}},
  \bibinfo{pages}{13544} (\bibinfo{year}{1992}).

\bibitem[{\citenamefont{Sonin}(2010)}]{sonin}
\bibinfo{author}{\bibfnamefont{E.~B.} \bibnamefont{Sonin}},
  \bibinfo{journal}{Adv. Phys.} \textbf{\bibinfo{volume}{59}},
  \bibinfo{pages}{181} (\bibinfo{year}{2010}).

\bibitem[{\citenamefont{Chen and Sigrist}(2014)}]{sigrist}
\bibinfo{author}{\bibfnamefont{W.}~\bibnamefont{Chen}} \bibnamefont{and}
  \bibinfo{author}{\bibfnamefont{M.}~\bibnamefont{Sigrist}},
  \bibinfo{journal}{Phys. Rev. B} \textbf{\bibinfo{volume}{89}},
  \bibinfo{pages}{024511} (\bibinfo{year}{2014}).

\bibitem[{\citenamefont{Brataas et~al.}(2002)\citenamefont{Brataas,
  Tserkovnyak, Bauer, and Halperin}}]{battery}
\bibinfo{author}{\bibfnamefont{A.}~\bibnamefont{Brataas}},
  \bibinfo{author}{\bibfnamefont{Y.}~\bibnamefont{Tserkovnyak}},
  \bibinfo{author}{\bibfnamefont{G.~E.~W.} \bibnamefont{Bauer}},
  \bibnamefont{and} \bibinfo{author}{\bibfnamefont{B.~I.}
  \bibnamefont{Halperin}}, \bibinfo{journal}{Phys. Rev. B}
  \textbf{\bibinfo{volume}{66}}, \bibinfo{pages}{060404(R)}
  (\bibinfo{year}{2002}).

\bibitem[{\citenamefont{Uchida et~al.}(2010)\citenamefont{Uchida, Xiao, Adachi,
  Ohe, Takahashi, Ieda, Ota, Kajiwara, Umezawa, Kawai
  et~al.}}]{uchidainsulator}
\bibinfo{author}{\bibfnamefont{K.}~\bibnamefont{Uchida}},
  \bibinfo{author}{\bibfnamefont{J.}~\bibnamefont{Xiao}},
  \bibinfo{author}{\bibfnamefont{H.}~\bibnamefont{Adachi}},
  \bibinfo{author}{\bibfnamefont{J.}~\bibnamefont{Ohe}},
  \bibinfo{author}{\bibfnamefont{S.}~\bibnamefont{Takahashi}},
  \bibinfo{author}{\bibfnamefont{J.}~\bibnamefont{Ieda}},
  \bibinfo{author}{\bibfnamefont{T.}~\bibnamefont{Ota}},
  \bibinfo{author}{\bibfnamefont{Y.}~\bibnamefont{Kajiwara}},
  \bibinfo{author}{\bibfnamefont{H.}~\bibnamefont{Umezawa}},
  \bibinfo{author}{\bibfnamefont{H.}~\bibnamefont{Kawai}},
  \bibnamefont{et~al.}, \bibinfo{journal}{Nat. Mater.}
  \textbf{\bibinfo{volume}{9}}, \bibinfo{pages}{894} (\bibinfo{year}{2010}).

\bibitem[{\citenamefont{Adachi et~al.}(2011)\citenamefont{Adachi, Ohe,
  Takahashi, and Maekawa}}]{adachi}
\bibinfo{author}{\bibfnamefont{H.}~\bibnamefont{Adachi}},
  \bibinfo{author}{\bibfnamefont{J.}~\bibnamefont{Ohe}},
  \bibinfo{author}{\bibfnamefont{S.}~\bibnamefont{Takahashi}},
  \bibnamefont{and} \bibinfo{author}{\bibfnamefont{S.}~\bibnamefont{Maekawa}},
  \bibinfo{journal}{Phys. Rev. B} \textbf{\bibinfo{volume}{83}},
  \bibinfo{pages}{094410} (\bibinfo{year}{2011}).

\bibitem[{\citenamefont{Holstein and Primakoff}(1940)}]{HP}
\bibinfo{author}{\bibfnamefont{T.}~\bibnamefont{Holstein}} \bibnamefont{and}
  \bibinfo{author}{\bibfnamefont{H.}~\bibnamefont{Primakoff}},
  \bibinfo{journal}{Phys. Rev.} \textbf{\bibinfo{volume}{58}},
  \bibinfo{pages}{1098} (\bibinfo{year}{1940}).

\bibitem[{\citenamefont{Takayoshi et~al.}(2014)\citenamefont{Takayoshi, Aoki,
  and Oka}}]{takayoshi}
\bibinfo{author}{\bibfnamefont{S.}~\bibnamefont{Takayoshi}},
  \bibinfo{author}{\bibfnamefont{H.}~\bibnamefont{Aoki}}, \bibnamefont{and}
  \bibinfo{author}{\bibfnamefont{T.}~\bibnamefont{Oka}},
  \bibinfo{journal}{Phys. Rev. B} \textbf{\bibinfo{volume}{90}},
  \bibinfo{pages}{085150} (\bibinfo{year}{2014}).

\bibitem[{\citenamefont{Nakata}(2012)}]{QSP}
\bibinfo{author}{\bibfnamefont{K.}~\bibnamefont{Nakata}}, \bibinfo{journal}{J.
  Phys. Soc. Jpn.} \textbf{\bibinfo{volume}{81}}, \bibinfo{pages}{064717}
  (\bibinfo{year}{2012}).

\bibitem[{\citenamefont{Rammer}(2007)}]{rammer}
\bibinfo{author}{\bibfnamefont{J.}~\bibnamefont{Rammer}},
  \emph{\bibinfo{title}{Quantum Field Theory of Non-equilibrium States}}
  (\bibinfo{publisher}{Cambridge University Press, Cambridge},
  \bibinfo{year}{2007}).

\bibitem[{\citenamefont{Tatara et~al.}(2008)\citenamefont{Tatara, Kohno, and
  Shibata}}]{tatara}
\bibinfo{author}{\bibfnamefont{G.}~\bibnamefont{Tatara}},
  \bibinfo{author}{\bibfnamefont{H.}~\bibnamefont{Kohno}}, \bibnamefont{and}
  \bibinfo{author}{\bibfnamefont{J.}~\bibnamefont{Shibata}},
  \bibinfo{journal}{Physics Report} \textbf{\bibinfo{volume}{468}},
  \bibinfo{pages}{213} (\bibinfo{year}{2008}).

\bibitem[{\citenamefont{Loss and Goldbart}(1996)}]{dipole}
\bibinfo{author}{\bibfnamefont{D.}~\bibnamefont{Loss}} \bibnamefont{and}
  \bibinfo{author}{\bibfnamefont{P.~M.} \bibnamefont{Goldbart}},
  \bibinfo{journal}{Phys. Lett. A} \textbf{\bibinfo{volume}{215}},
  \bibinfo{pages}{197} (\bibinfo{year}{1996}).

\bibitem[{\citenamefont{Caldeira and Leggett}(1983)}]{C-L}
\bibinfo{author}{\bibfnamefont{A.~O.} \bibnamefont{Caldeira}} \bibnamefont{and}
  \bibinfo{author}{\bibfnamefont{A.~J.} \bibnamefont{Leggett}},
  \bibinfo{journal}{Physica A} \textbf{\bibinfo{volume}{121}},
  \bibinfo{pages}{587} (\bibinfo{year}{1983}).

\bibitem[{\citenamefont{Breuer and Petruccione}(2002)}]{CL_text}
\bibinfo{author}{\bibfnamefont{H.-P.} \bibnamefont{Breuer}} \bibnamefont{and}
  \bibinfo{author}{\bibfnamefont{F.}~\bibnamefont{Petruccione}},
  \emph{\bibinfo{title}{The Theory of Open Quantum Systems}}
  (\bibinfo{publisher}{Cambridge University Press, Oxford},
  \bibinfo{year}{2002}).

\bibitem[{\citenamefont{Nikuni et~al.}(2000)\citenamefont{Nikuni, Oshikawa,
  Oosawa, and Tanaka}}]{oshikawa}
\bibinfo{author}{\bibfnamefont{T.}~\bibnamefont{Nikuni}},
  \bibinfo{author}{\bibfnamefont{M.}~\bibnamefont{Oshikawa}},
  \bibinfo{author}{\bibfnamefont{A.}~\bibnamefont{Oosawa}}, \bibnamefont{and}
  \bibinfo{author}{\bibfnamefont{H.}~\bibnamefont{Tanaka}},
  \bibinfo{journal}{Phys. Rev. Lett.} \textbf{\bibinfo{volume}{84}},
  \bibinfo{pages}{5868} (\bibinfo{year}{2000}).

\bibitem[{\citenamefont{Rezende}(2009)}]{rezende}
\bibinfo{author}{\bibfnamefont{S.}~\bibnamefont{Rezende}},
  \bibinfo{journal}{Phys. Rev. B} \textbf{\bibinfo{volume}{79}},
  \bibinfo{pages}{174411} (\bibinfo{year}{2009}).

\bibitem[{\citenamefont{Troncoso and $\acute{\rm{A}}$.
  S.~N$\acute{\rm{u}}$$\tilde{\rm{n}}$ez}(2012)}]{troncoso}
\bibinfo{author}{\bibfnamefont{R.~E.} \bibnamefont{Troncoso}} \bibnamefont{and}
  \bibinfo{author}{\bibnamefont{$\acute{\rm{A}}$.
  S.~N$\acute{\rm{u}}$$\tilde{\rm{n}}$ez}}, \bibinfo{journal}{J. Phys.:
  Condens. Matter} \textbf{\bibinfo{volume}{24}}, \bibinfo{pages}{036006}
  (\bibinfo{year}{2012}).

\end{thebibliography}

\end{document}